\documentclass[%
 aip,
 rsi,
 amsmath,amssymb,
reprint,%
groupedaddress, 
]{revtex4-1}

\usepackage{graphicx}
\usepackage{dcolumn}
\usepackage{bm}

\usepackage[utf8]{inputenc}
\usepackage[T1]{fontenc}
\usepackage{mathptmx}
\usepackage{etoolbox}
\usepackage{adjustbox}

\def\0{0}

\def\sam{\textrm{sam}}

\def\PMT{\textrm{PMT}}

\makeatletter
\def\@email#1#2{%
 \endgroup
 \patchcmd{\titleblock@produce}
  {\frontmatter@RRAPformat}
  {\frontmatter@RRAPformat{\produce@RRAP{*#1\href{mailto:#2}{#2}}}\frontmatter@RRAPformat}
  {}{}
}%
\makeatother
\begin{document}


\title[]{Spectroradiometry with sub-microsecond time resolution using multianode photomultiplier tube assemblies}
\author{Zachary M. Geballe}
\author{Francesca Miozzi}
\author{Chris F. Anto}
 \altaffiliation[Also at ]{Department of Physics and Astronomy, Johns Hopkins University.}
\author{Javier Rojas}
\author{Jing Yang}
\author{Michael J. Walter}
\affiliation{ 
Earth and Planets Laboratory, Carnegie Institution for Science
}%

\date{\today}

\begin{abstract}
Accurate and precise measurements of spectroradiometric temperature are crucial for many high pressure experiments that use diamond anvil cells or shock waves. In experiments with sub-millisecond timescales, specialized detectors such as streak cameras or photomultiplier tubes are required to measure temperature. High accuracy and precision are difficult to attain, especially at temperatures below 3000 K.  Here we present a new spectroradiometry system based on multianode photomultiplier tube technology and passive readout circuitry that yields a 0.24~$\mu$s rise-time for each channel. Temperature is measured using five color spectroradiometry. During high pressure pulsed Joule heating experiments in a diamond anvil cell, we document measurement precision to be~$\pm 30$~K at temperatures as low as 2000 K during single-shot heating experiments with~$0.6$~$\mu$s time-resolution. Ambient pressure melting tests using pulsed Joule heating indicate that the accuracy is~$\pm 80$~K in the temperature range 1800 -- 2700 K.
\end{abstract}

\maketitle

\section{Introduction}
Pulsed heating in diamond anvil cells (DACs) is useful for measuring thermal conduction,\cite{Yagi2011,McWilliams2015} synthesizing new materials,\cite{Peiris2003,Dubrovinsky2022,Mishra2018} creating exotic high density fluids,\cite{Zaghoo2016,McWilliams2016} and detecting the latent heat of melting.\cite{Geballe2021} 
The most common way to measure sample temperatures in pulsed heated DAC experiments is by fitting a sample’s thermal emissions spectra to a greybody function, a method known as ``spectroradiometry’’.  Together, spectroradiometry and pulsed heating in DACs have enabled several breakthrough measurements. They were used to document the insulating-to-conducting transition in fluid hydrogen,\cite{Zaghoo2016,Zaghoo2018,McWilliams2016} and to measure thermal conductivity of silicates and metals at the pressure-temperature conditions of Earth's lower mantle and outer core. \cite{Geballe2021,Konopkova2016} Spectroradiometry has also been used to determine temperature during shock wave experiments.\cite{Boslough1989,Asimow2015} 

Beyond the experimental challenges in all spectroradiometry measurements,\cite{Benedetti2004,Benedetti2007,Walter2004} two additional challenges arise because of the short duration of pulsed heating experiments and shock wave experiments: the measuring device requires (a) fast time resolution and (b) low enough noise for precise temperature fits. The desired precision is typically in the range of 10 to 100 K. The desired time resolution varies from ns to ms.\cite{Zaghoo2016, McWilliams2015, Zhang2015, Boslough1989}

The ideal spectroradiometry system would enable high precision measurements at a wide range of temperatures with high time resolution during single shot experiments (i.e., without need to accumulate signal during multiple experiments). In practice, tradeoffs are necessary. Several studies have shown measurements at temperatures as low as~$\sim 2000$~K with $\mu$s to ms time resolution while accumulating thermal emissions for 10s to 100s of ms.\cite{Zhang2015, McWilliams2015, Beck2007} Ref. \onlinecite{Montgomery2018} extended the low temperature limit using 1 ms time resolution, reaching 1300 K during single shot experiments, and temperature as low as 800 K during during multi-shot experiments. Refs. \onlinecite{McWilliams2015PNAS,Geballe2021} extended the lower limit of cumulative heating duration. They used streak cameras to measure temperatures in the range 3000 K to 13,000 K with~$\sim 0.5$~$\mu$s time resolution while accumulating data with 1 to 10 repetitions.

In the present study, we extend the lower limit of temperature far below 3000 K while maintaining sub-$\mu$s time resolution during single shot experiments. Notably, this opens up the possibility of accurate temperature measurements at conditions typical of Earth's lower mantle ($\sim 1800 - 2500$~K and ~$23 - 136$~GPa) while performing single shot, $\mu$s-timescale heating experiments in DACs. We also improve temperature precision relative to our previous work.\cite{Geballe2021} Moreover, the cost of the system is reduced substantially compared to optical systems based on expensive streak cameras. The key technology is a multianode photomultiplier tube (MA-PMT) assembly. MA-PMT assemblies have been used for spectroscopic measurements of chemical kinetics during shock experiments.\cite{Bouyer2006,Matsugi2016,Matsugi2020} They are a new technology for spectroradiometry in DACs.

One MA-PMT assembly is installed as the optical detector for spectroradiometry on each side of a new optical table at the Carnegie Earth and Planets Laboratory. The table is designed for pulsed Joule heating and spectroradiometry of diamond anvil cell samples. In section II, we present details of the new optical table. In section III, we show the results of calibrations and tests using incandescent lamps and pulsed LEDs. In section IV, we describe the data processing method. Section V shows examples of heating at high pressure ($\sim 76$~GPa). Lastly, section VI assesses the accuracy of the temperature measurement based on melting tests on four ambient-pressure samples: iron, platinum, alumina, and iridium. 

\section{Optical and electronic setup}
We divide the description of the new optical table into three parts: the optics, the MA-PMT assemblies, and the readout electronics.

\begin{figure*}[tbhp]
	\centering
	\includegraphics[width=.9\linewidth]{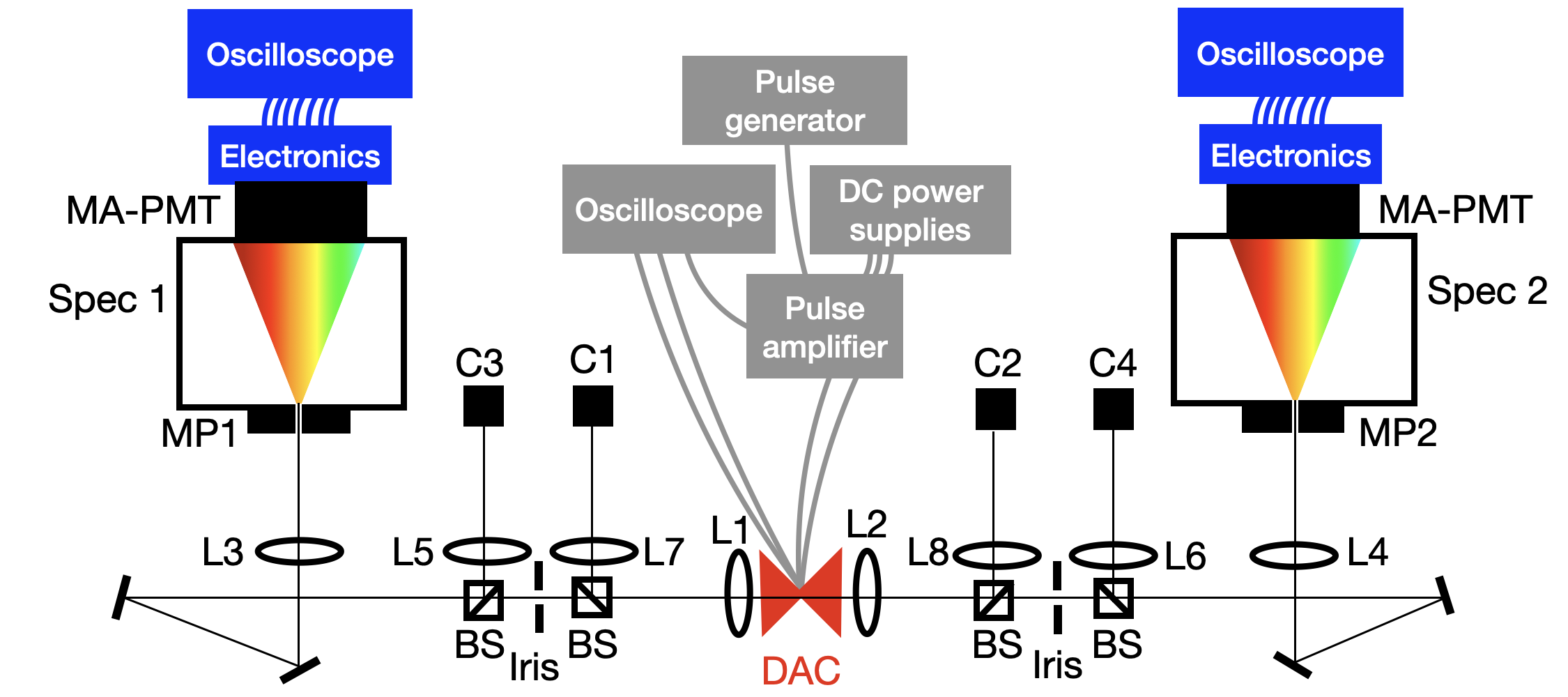}
	\caption{Schematic of the optical table. This paper focuses on the optics (black) and the readout electronics for the MA-PMTs (blue). Grey shading indicates the electronics for pulsed Joule heating; see Ref. \onlinecite{Geballe2023Pulser} for details. Objective lenses L1 and L2 (20X Mitutoyo Plan Apo NIR Infinity Corrected) collimate thermal emissions from the diamond anvil cell (DAC). Ten percent of the collimated light is reflected by cube beamsplitters (BS; ThorLabs BS025) and focused by achromatic lenses (L7, L8; ThorLabs AC254-200-AB) onto CMOS cameras (C1, C2; FLIR BFS-PGE-16S2C-CS Blackfly S PoE, Color). The transmitted light passes through irises (Iris; ThorLabs SM1D12C) and cube beamsplitters (same part as above), and is focused by achromatic focusing lenses (L3, L4; ThorLabs AC254-200-AB) onto the mirror pinholes (MP1, MP2; aluminum coated Edmund Optics 45-658) located at the focal positions of the two spectrometers. The images of the sample on the mirror pinholes are reflected back along the incident beam paths to the outer beamsplitters. Ten percent of the mirror-pinhole-reflected image is reflected by cube beamsplitters, focused by achromatic lenses (L5, L8; same part as above) and captured by CMOS cameras (C3, C4; BFS-U3-16S2C-CS USB 3.1 Blackfly S, Color). The light that passes through mirror pinhole is chromatically dispersed by the spectrometers (Holospec VPH with HFG-750 transmission grating on the left side; Acton  SP2358 with 150 g/mm grating on the right side). The chromatically dispersed light (shown schematically as a rainbow), is detected by MA-PMT assemblies (Hamamatsu H7260-20). Blue boxes and curves indicate electronics and electrical connections. The schematic does not show the following optical elements that are removed from the beam path during heating experiments: shutters (ThorLabs SH05) are located between the focusing lenses and the mirror pinholes; white LED illumination is delivered by pellicle beamsplitters (ThorLabs BP145B1) on electrical flipper mounts (New Focus 8893) located between the objective lenses and cube beamsplitters.}
	\label{fig:optics_schematic}
\end{figure*}

\begin{figure*}[tbhp]
	\centering
	\includegraphics[width=.9\linewidth]{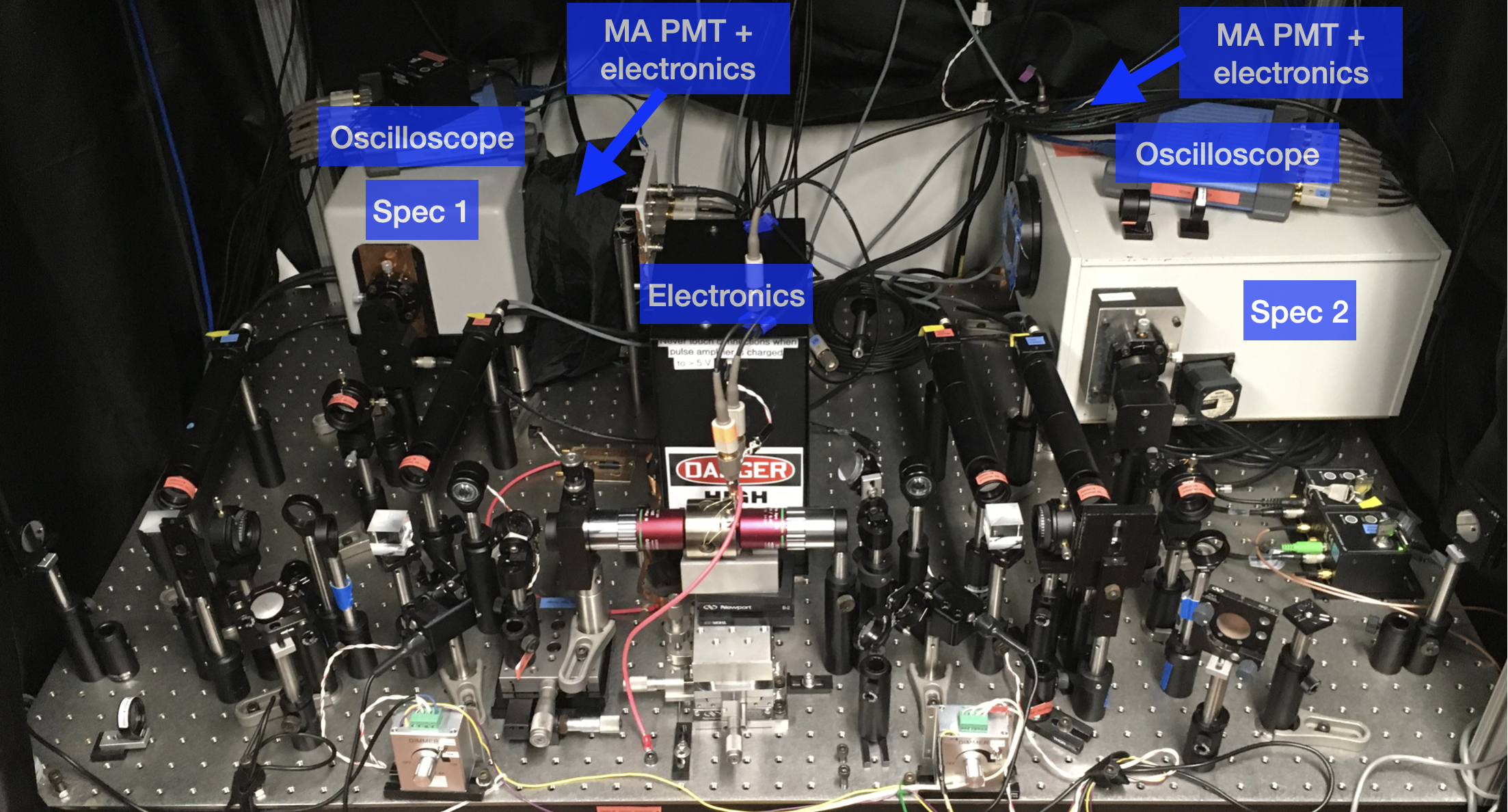}
	\caption{Photograph of the optical table, with blue labels marking the spectrometers, MA-PMT assemblies, and electronics including oscilloscopes.}
	\label{fig:table_photo}
\end{figure*}

\subsection{Optics}
The optical table design is shown schematically in Fig. \ref{fig:optics_schematic} and in a photo in Fig. \ref{fig:table_photo}. It borrows many design attributes and optical component choices from existing laser heating systems for diamond anvil cells (DACs),\cite{Prakapenka2008, Meng2015, McWilliams2015, Kantor2018} including the following: apochromatic objective lenses to collimate the sample’s thermal emissions, irises to reduce chromatic aberration, and mirror pinholes at the front of a spectrometer. A CMOS camera monitors the position of the sample image on the mirror pinhole. 

Several details of the optical system are different from all eight of the laser heating systems described in \onlinecite{Prakapenka2008,Meng2015,McWilliams2015,Kantor2018,Petitgirard2014,Anzellini2018,Konopkova2021,Watanuki2001}.  First, each mirror pinhole was fabricated in-house using an RF-magnetron sputtering system to deposit a~$\sim 400$~nm thick aluminum film onto the surface of an anti-reflective coated piece of optical glass (Edmund Optics 45-658). Before the sputter-deposition, a 100~$\mu$m diameter disc of platinum was placed at the center of the optical glass as a mask. After sputter-deposition, the mask was removed. Appendix \ref{SM:mirror_pinhole} describes masking and sputtering details. Second, the main focusing lenses (L3 and L4 in Fig. \ref{fig:optics_schematic}) perform two functions: they focus thermal emissions onto the mirror pinhole, and they collimate light reflected from the mirror pinhole. This design feature saves space on the optical table. Third, we include a second CMOS camera on each side (C1,C2), along with a stationary beamsplitter and a lens (L7,L8) to monitor the entire sample, including the 5~$\mu$m-diameter disc that images onto the pinhole. The images from CMOS cameras C1 and C2 during heating are useful for estimating the spatial distribution of sample temperature using pyrometry (e.g. Fig. S1 of Ref. \onlinecite{Geballe2020}). We use cube beamsplitters rather than pellicle beamsplitters because their transmission spectra are much smoother, and because they are less prone to mechanical damage. The downside of cube beamsplitters is that, despite their anti-reflective coating, they can create ghost images that can focus onto the mirror pinhole. In the present setup, ghost images were avoided by rotating each cube beamsplitter~$3^\circ$~away from the optical axis, causing reflections from the cube surfaces to avoid all the nearby reflective surfaces on our table. 

The two spectrometers were repurposed from previous experimental setups. The left side uses a Holospec VPH spectrometer equipped with a HFG-750 transmission grating. The right side uses an Acton SP2358 spectrometer equipped with a 150 g/mm grating blazed at 500 nm. A more symmetric design with a smaller footprint would use 150 mm focal length reflecting spectrometers with 300 g/mm grating on each side. 

\subsection{MA-PMT assemblies}
Thermal emissions are detected on each side of the DAC sample using a multi-anode photomultiplier tube (MA-PMT) assembly (Hamamatsu H7260-20). Each MA-PMT assembly consist of a linear array of 32 photomultiplier tubes with hard-wired voltage dividing circuitry. Similar MA-PMT assemblies are used in the shock-wave spectroscopy systems of Refs. \onlinecite{Bouyer2006,Matsugi2016}. In our setup, the package is housed in a 10x10x3 cm aluminum holder that is attached to either the spectrometer itself (for the Acton spectrometer) or to a motorized linear translation stage positioned in front of the spectrometer output (for the Holospec spectrometer).

Each MA-PMT assembly is powered by a high voltage power supply (SRS PS325 and SRS PS310  for left and right sides, respectively). In practice, we operate the power supplies in the range -500 V to -900 V, which corresponds to gains in the range~$10^4$~to~$3 \times 10^6$. 

\begin{figure}[tbhp]
	\centering
	\includegraphics[width=3.3in]{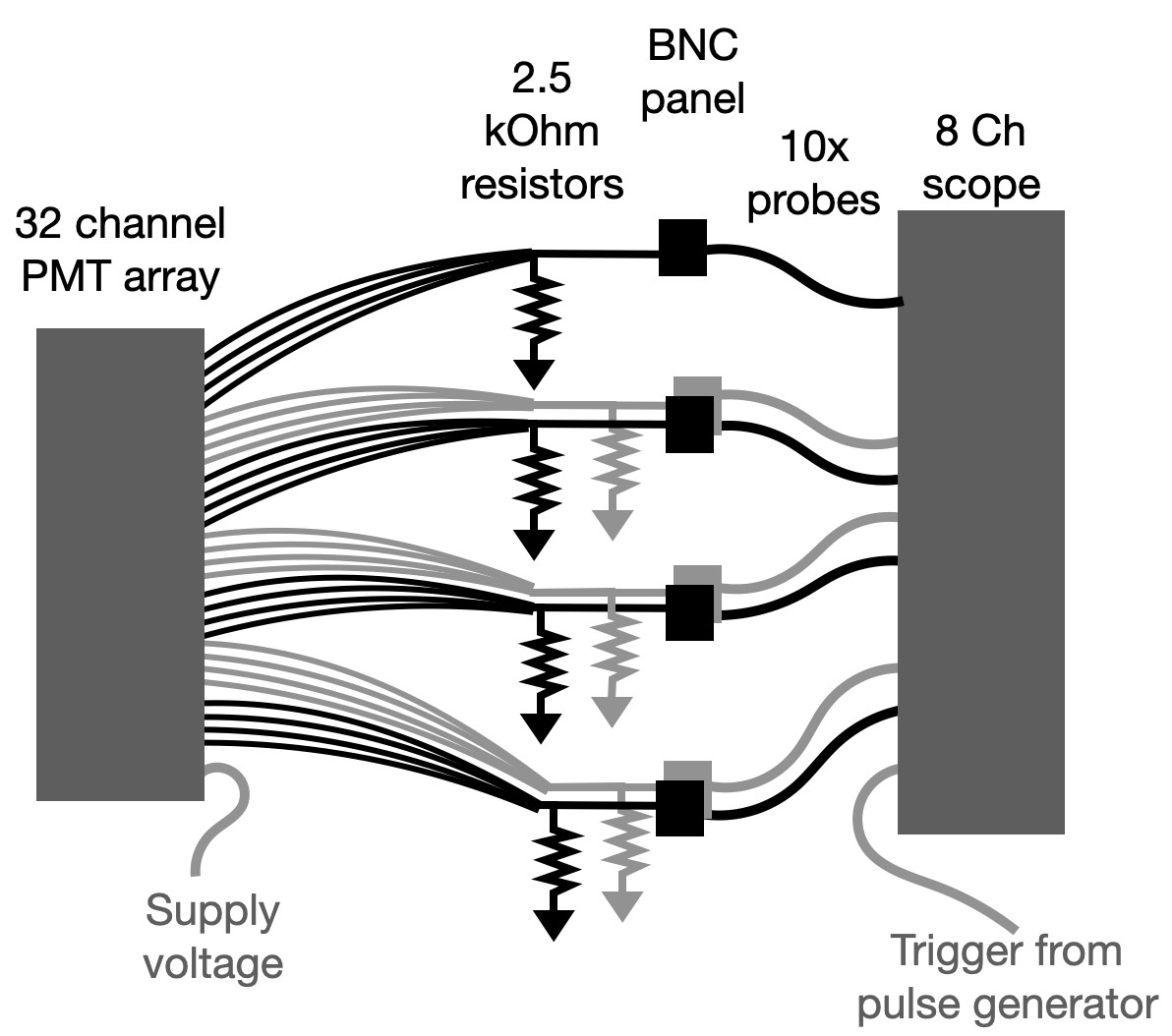}
	\caption{Schematic of the MA-PMT assembly and readout electronics used on the right side of the optical table. Counterclockwise from bottom left: a supply voltage from a high voltage power supply (SRS PS310) sets the gain on the 32-channel MA-PMT assembly (Hamamatsu H7260-20). Seven sets of four PMT anodes are electrically connected. Each set of anodes is shunted to ground through a 2.5 k$\Omega$~resistors (1\% accuracy, 1/8 W), and connected via a BNC panel mount to a 10x oscilloscope probe (ProbeMaster 6143-4), which is connected to one channel of an 8-channel oscilloscope (Picoscope 4824A; 20 MHz, 12-bit). One channel of the oscilloscope is used to received the trigger from a pulse generator.}
	\label{fig:electrical_schematic}
\end{figure}

\subsection{MA-PMT readout electronics}
The outputs of the 32 MA-PMT channels are bunched into 6 groups of 5 for the Holospec spectrometer on the left side of the optical table, and 7 groups 4 for the Acton spectrometer on the right side of the optical table. Each group of outputs is combined electrically and shunted to ground through a 2.5 k$\Omega$~resistor, as shown schematically in Fig. \ref{fig:electrical_schematic}. The unused PMT outputs are grounded. Output voltage is measured by 10x oscilloscope probes connected to 8-channel oscilloscopes.
The choice of 2.5 k$\Omega$~resistance is low enough to yield sub-$\mu$s time resolution and high enough so that~$\sim 2000$~K metal samples produce voltages in most oscilloscope channels that are several orders of magnitude above the 2~$\mu$V quantization noise of the oscilloscope. (The exceptions are the bluest channel and farthest infrared channel, which usually do not detect signal above the noise, leaving five usable oscilloscope channels out of seven). 

\section{Calibration and MA-PMT characterization}

The MA-PMT assemblies are the major difference in this optical table compared to previous spectroradiometry systems for pulsed laser heating and pulsed Joule heating.\cite{McWilliams2015, Meng2015,Beck2007, Prakapenka2008} Four key challenges in using a MA-PMT assembly are: (1) to ensure data is collected in the linear regime of the MA-PMT assembly, (2) to ensure a desirable combination of time resolution, wavelength resolution, and noise floor, (3) to calibrate the wavelength range of each readout channel, and (4) to apply sufficient grounding so that MA-PMTs and their readout electronics are negligibly affected by the Joule heating pulse - a challenge that is specific to pulsed Joule heating. Below, we describe the results of tests for each of these challenges, plus a  system response measured with a NIST-traceable tungsten lamp to calibrate the spectroradiometric measurements.

\begin{figure}[tbhp]
	\centering
	\includegraphics[width=.9\linewidth]{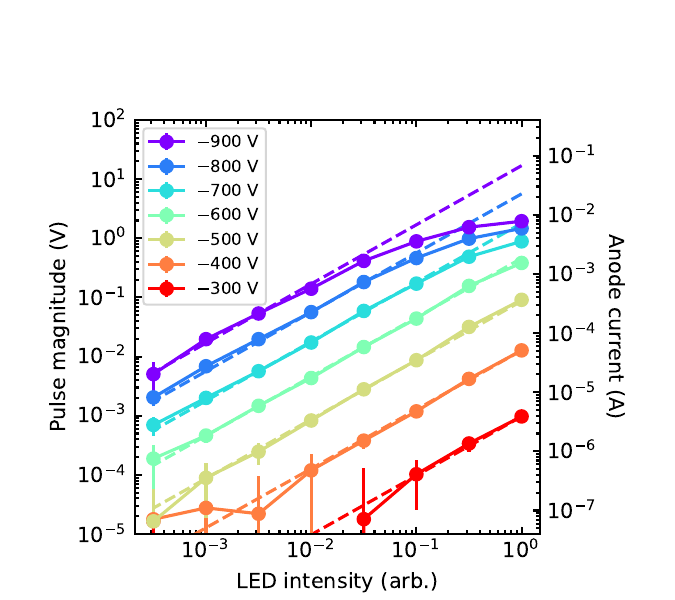}
	\caption{Linearity testing results. Circles and solid curves show the amplitude of the PMT measurement of a 20~$\mu$s pulse from a high-powered green LED (Luminus SBT-70-G) on the right side of the optical table using variable gain (different colors), and variable attenuation (horizontal axis). Pulse amplitude is recorded from the channel 4 of the oscilloscope, which measures wavelength in the range 530-600 nm. Dashed lines show a linear response (i.e. slope = 1 in log-log space). The legend lists the voltages used to set the gain of the MA-PMT assembly.}
	\label{fig:linearity_test}
\end{figure}

\subsection{MA-PMT Linearity}
\label{section:linearity}
The range of anode currents over which a PMT is linear depends strongly on the pulse duration.\footnote{See www.hamamatsu.com/content/dam/hamamatsu-photonics/sites/documents/ 99\_SALES\_LIBRARY/etd/PMT\_handbook\_v4E.pdf} \cite{Bassett2016} Our system is designed for pulse durations in the range~$\sim 1$~to 100~$\mu$s. Fig. \ref{fig:linearity_test} shows that at 20~$\mu$s pulse duration, one of the central channels of the detector is approximately linear up to~$\sim 0.4$~mA (100 mV) at -700 to -900 V supply voltages. The data at lower supply voltages remains linear up to the highest light levels tested. The anode current is linear down to at least 0.1~$\mu$A at -500 V supply voltage. Our measurements at lower anode currents are not precise enough to test linearity. The~$\sim 0.1$~$\mu$A - 0.4 mA range documented here is meant only as a guide, because the effect of non-linearity on spectroradiometry is a complicated function of temperature-time history and supply voltage. Below, we use a different test to show that anode non-linearity does not affect our measured temperatures during Joule heating experiments. We vary supply voltage in the range -550 to -800 V (hence varying anode current 30-fold) and show that the change in fitted temperature is within the scatter of temperature measurements. Appendix \ref{SM:PMT_linearity} describes how the choice of MA-PMT assembly and its optical focusing procedure may affect linearity.  

\begin{figure}[tbhp]
	\centering
	\includegraphics[width=3in]{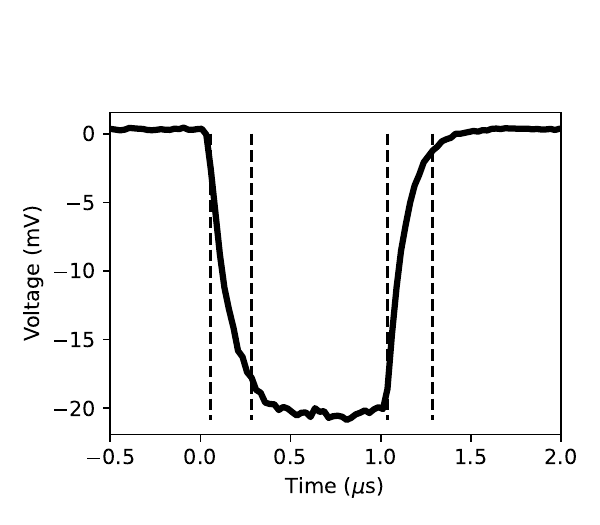}
	\caption{Time resolution. The black curve is channel 3 (620-690 nm) of the right side of the optical table, measured while pulsing a red LED (ThorLabs LED630L) using a 1 $\mu$s square wave output from a pulse generator (Berkeley Nucleonics 525). Vertical dashed lines mark 10\% and 90\% of the full rise and fall. The rise and fall times are 0.23~$\mu$s and 0.25~$\mu$s.}
	\label{fig:timing_test}
\end{figure}

\subsection{Time resolution}
We measure the time resolution of our detector by measuring the response to a red LED that is driven by a 1~$\mu$s square wave from a pulse generator. Fig. \ref{fig:timing_test} shows that the typical rise time and fall time is 0.24~$\pm 0.1$~$\mu$s. Here we define rise and fall times by the times that the signal reaches 10 and 90\% of the pulse magnitude, a definition that corresponds to $2.2 R C$ for an ideal circuit made up of a single resistor with resistance $R$ and a single capacitor with capacitance $C$. This suggests that the total capacitance per oscilloscope channel is~$C = 0.24$~$\mu$s$/2.2/2.5$k$\Omega$~$ = 40$~pF. (The total capacitance is the sum each electrical lead self capacitance and the output capacitances of four PMT anodes). Further testing with shunt resistances in the range 1-10 k$\Omega$~confirms that the rise time and fall time of the detector varies in proportion to the value of shunt resistance.

\begin{figure}[b!]
	\centering
	\includegraphics[width=3.5in]{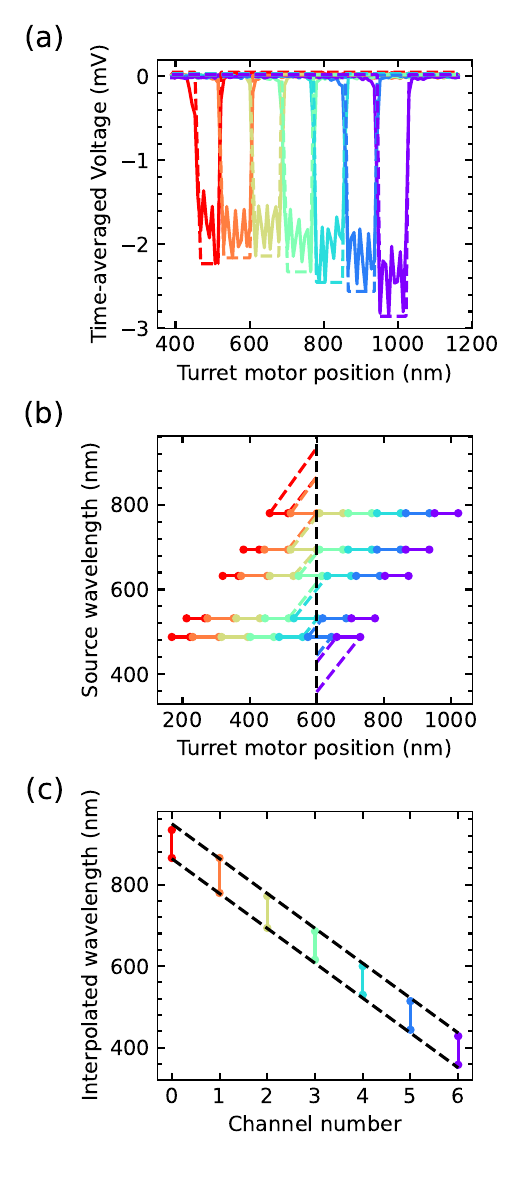}
\end{figure}
\begin{figure}[t!]
	\centering
	\caption{Wavelength calibration for the right side spectrometer and MA-PMT assembly. (a) Time-averaged voltages (solid curves) and best-fit square wave (dashed curves) for each oscilloscope channel (colors) when the mirror pinhole is illuminated by~$780 \pm 5$~nm light, plotted as a function of the motor position of the spectrometer's turret. (b) A summary of the square-waves fitted to data from (a) and the analogous data for 488, 532, 632, and 694 nm light. Circles and line segments mark the ranges where fitted square waves are less than 0. Dashed colors mark linear interpolations and extrapolations of the circles. The motor position used in experiments is marked by the vertical dashed black line at 600 nm. (c) The locations of linear interpolations and extrapolations to 600 nm motor position from (b) are marked as colored circles. Black dashed lines show the final wavelength calibration, a linear fit to channels 1-6.}
	\label{fig:wavelength_calibration}
\end{figure}

\subsection{Wavelength calibration}
Wavelength calibration with a MA-PMT assembly presents a subtle challenge. A calibration is required because the relative alignment of detector and spectrometer affects the wavelengths detected on each channel. (Note that this is unlike previous DAC studies such as Refs. \onlinecite{Montgomery2018,Zhang2015} in which notch filters are used to separate colors). And yet there are not enough channels to use a high spectral resolution standard source like a neon lamp that is typical when calibrating a CCD detector. Finally, we do not directly measure the wavelength range detected by each channel because we do not own continuously tunable monochromatic light sources (e.g. a set of tunable dye lasers). 

Rather than calibrating channels directly, we use a three-step analysis procedure based on intensity data collected as a function of motor position, where the motor scans the MA-PMT assembly with respect to the location of a narrow bandwidth of light. First, we illuminate the pinhole of the spectrometer with a 10 nm band of light (e.g.~$532 \pm 5$~nm) by backlighting a 10 nm bandpass filter with an incandescent lamp. We then use a motor to scan the position of the MA-PMT assembly relative to the narrow band of light. On the Holospec side, we use a linear motor to translate the MA-PMT assembly. On the Acton side, we use the spectrometer’s turret motor to move the narrow band of light. An example of the resulting intensity scan on the Acton side is shown in Fig. \ref{fig:wavelength_calibration}a. The first analysis step is to identify the range of motor positions for which the light is detected by each channel of the MA-PMT detector, ignoring any channels that are not fully transited by the motor scan. Here, ``detected’’ means that the measured voltage is at least half its maximum value during the scan. We perform this procedure for each of the following bands of light:~$488 \pm 5$~nm,~$532 \pm 5$~nm,~$632 \pm 5$~nm,~$694 \pm 5$~nm,~$780 \pm 5$~nm. Fig. \ref{fig:wavelength_calibration}b shows the range of motor positions plotted versus the central wavelength of the light source for each channel. The second analysis step is linear interpolation and extrapolation to the motor position that is used for experiments. We use the ``600 nm'' position of the Acton turret motor in our experiments. The interpolations and extrapolations are shown as dashed lines in Fig. \ref{fig:wavelength_calibration}b. Linear extrapolation is reasonable since the spectrometer grating dispersion is nearly linear, and since the photocathodes in the MA-PMT assembly are evenly spaced. Finally, we fit the central wavelengths of channels 1-6 of the Acton side and channels 1-4 of the Holospec side to a linear function. We use this linear function as our wavelength calibration. 

\begin{figure*}[tbhp]
	\centering
	\includegraphics[width=.9\linewidth]{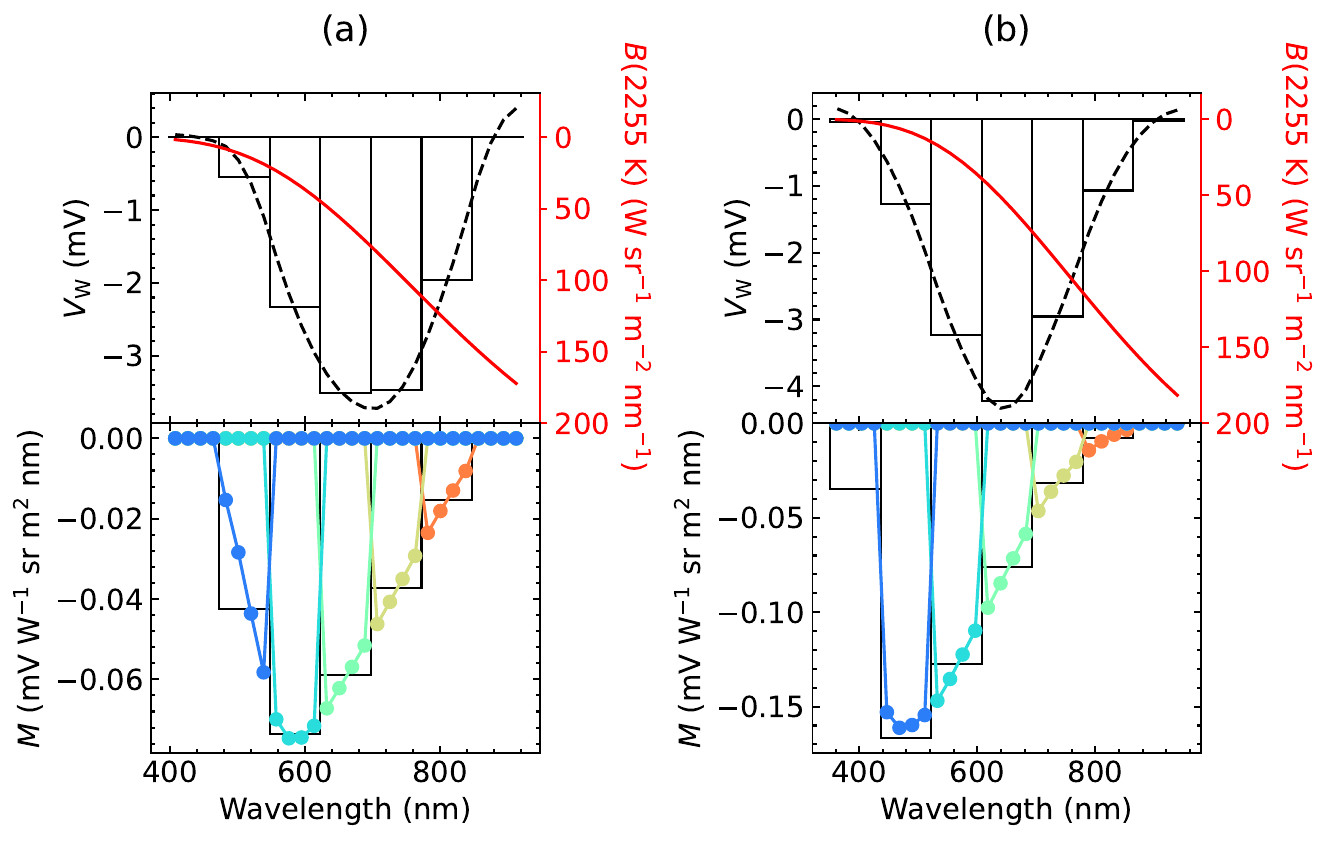}
	\caption{System response on left (a) and right (b) sides. (Top row) Time averaged voltage measured ($V_\textrm{W}$, solid black) and interpolated ($V_{W,\textrm{interp}}$, dashed black). The measurement uses a tungsten lamp heated to 2255 K color temperature and supply voltages -650 V and -700 V on left and right sides, respectively. The blackbody function~$B$(2255 K) is plotted in red. (Bottom row) A visual representation of the system response matrix,~$M = V_\textrm{W,interp}/B$. Each color corresponds to a row of the matrix. Each wavelength corresponds to a column of the matrix. In this figure, the oscilloscope channels are divided into 4 wavelengths for ease of visualization; 100 divisions per channel are used during analysis. For comparison, the black rectangles show the naive ratio calculated at the central wavelength,~$V_\textrm{W}(\lambda_i)/B$(2255 K,$\lambda_i)$.}
	\label{fig:QE}
\end{figure*}

\subsection{System response}
The ``system response’’ of each side of the optical table is the wavelength-dependent function that describes the voltage generated at the oscilloscope per photon emitted at the sample position. It is the wavelength-dependent transmission of all optics times the efficiency of the MA-PMT assembly and readout electronics in converting photons to volts. We follow the typical procedure for spectroradiometry, placing the filament of a NIST-traceable tungsten lamp at the sample position. We use 500 mA to heat the single-loop tungsten lamp described in Ref. \onlinecite{Karandikar_thesis} to the color temperature~$T=2255$~K. We set the supply voltage of the left and right MA-PMT assemblies to -650 and -700 V in order to achieve a sufficiently low anode current ($< 10$~$\mu$A) to avoid non-linearities, which occur around 15~$\mu$A when illuminating with continuous light. The resulting system response for left and right sides are plotted in Fig. \ref{fig:QE}a,b.


\begin{figure*}[tbhp]
    \centering
    \includegraphics[width=3in]{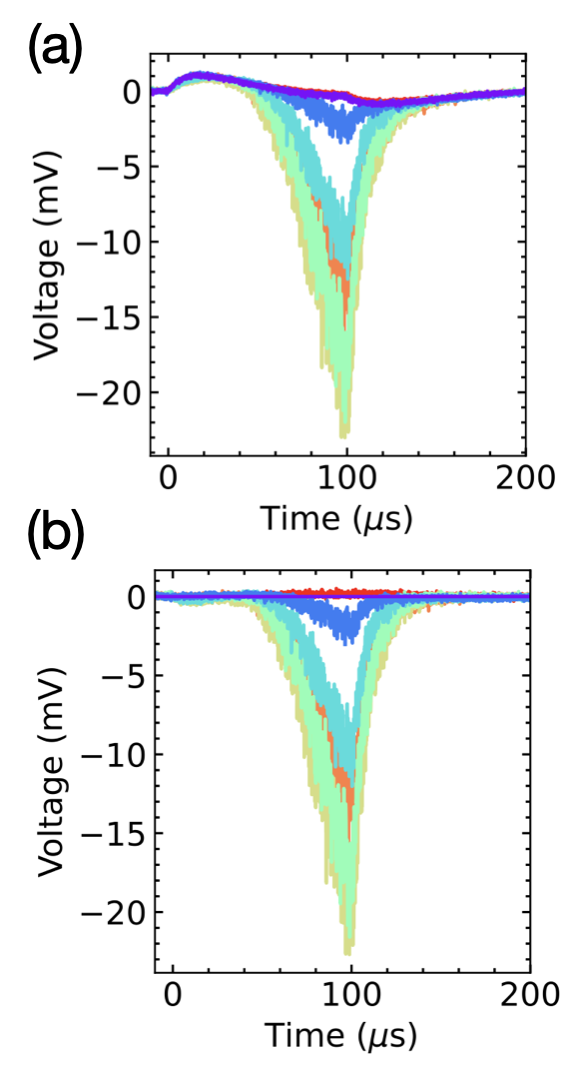}
    \caption{An example of the correction for electrical pickup in raw oscilloscope traces. (a) Raw traces for each channel on the left side of the optical table during pulsed heating of platinum to 2000 K using a current pulse that peaks at 10 A. The MA-PMT assembly supply voltage is -850 V. (b) Traces after subtracting the channel plotted in purple. The channel is not connected to the MA-PMT assembly, but rather measures spurious pickup induced through the 10x probe and 2.5 k$\Omega$~resistor. }
    \label{fig:electrical_pickup}
\end{figure*}

\subsection{Electrical pickup}
\label{section:pickup}
A spurious signal is generated in raw PMT voltages due to electrical pickup of large pulses of Joule heating current. For example, Fig. \ref{fig:electrical_pickup}a shows the result of a~$\sim 10$~A, 100~$\mu$s pulse of electricity through a platinum sample. The dominant signal in Fig. \ref{fig:electrical_pickup}  is caused by thermal emissions as the sample heats to~$\sim 2000$~K, a relatively low temperature. The spurious signal is the~$\sim 1$ mV oscillation with period~$\sim 200$~$\mu$s. The spurious signal scales with current amplitude, and occurs even when the optical shutter is closed or when the sample is replaced by a dummy resistor, indicating that it comes from electrical pickup. The spurious pickup signal is approximately independent of oscilloscope channel, allowing us to use an edge channel (channel 0 or 6) from each oscilloscope as the background (e.g. Fig. \ref{fig:electrical_pickup}b). (An alternative scheme for even better accuracy is to subtract background curves for each channel with the optical shutter closed). See Appendix \ref{SM:electrical_pickup} for further details.

\section{Data processing}

Time-resolved temperature is inferred using the following two-step method, which is similar to the methods in Refs. \onlinecite{Geballe2021,McWilliams2015}: 

\subsection{Step 1 of temperature analysis: a single two-parameter fit}

 A two parameter fit for temperature ($T$) and emissivity ($\epsilon$) is performed for voltage data averaged across a time interval of interest. In the example shown in Fig. \ref{fig:Fe_example}, we choose the time-interval 5.0 - 5.5~$\mu$s. The thermal emissions are averaged over the time-interval of interest and over all pulse repetitions, yielding a single spectrum,~$V_i$. Here,~$i$ denotes a channel of the oscilloscope, which corresponds to a band of wavelengths.  

The measured spectrum is fit to a greybody function multiplied by the system response matrix,~$M$, and integrated over wavelength:

\begin{equation}
V_i = \epsilon \int{ M(i,\lambda) B(T,\lambda) d\lambda}
\label{eqn:TwoParamPlanck}
\end{equation}
Here,~$B$ is Planck's blackbody function:
\begin{equation}
B = \frac{2 h c^2}{\lambda^5} \frac{1}{e^{\frac{hc}{\lambda k_B T}}-1} 
\end{equation}
where~$c$~is the speed of light,~$h$~is Planck's constant,~$k_B$~is Boltzmann's constant, and~$\lambda$ is wavelength. The system response matrix is defined by
\begin{eqnarray}
M(i,\lambda) &=& \frac{V_{\textrm{W,interp}}}{ B(\textrm{2555 K},\lambda)} \textrm{  if $\lambda_i - \delta\lambda/2<\lambda < \lambda_i + \delta\lambda/2$   } \\
&=& 0 \textrm{  otherwise}
\end{eqnarray}
Temperature ($T$) and emissivity ($\epsilon$) are the two free parameters. Variables~$\lambda_i$~and~$\delta\lambda$~denote the center and bandwidth of the measured wavelength band. The wavelength-dependent voltage,~$V_{\textrm{W,interp}}$, is an interpolation of the voltage,~$V_\textrm{W}$, that is measured during calibration with a standard tungsten lamp heated to a color temperature of 2255 K. See Appendix \ref{SM:W_interpolation} for interpolation details. The top panels in Figs. \ref{fig:QE}a,b show~$V_\textrm{W}$ (solid black),~$V_\textrm{W,interp}$ (dashed black), and~$B$(2255 K) (red), while the bottom shows the system response matrix,~$M$. Note that interpolation is required to achieve accuracy~$< 50$~K in this study, because the bandwidth of each oscilloscope channel is~$\sim 80$~nm. For comparison, spectroradiometry using a CCD or streak camera does not require interpolation because intensities are measured over narrow wavelength bands; a single pixel typically detects~$\sim 0.5$~nm. In those cases, the system response matrix is replaced by the normal system response. As another comparison, Ref. \onlinecite{Zhang2015} uses a different type of interpolation to account for the~$\sim 40$~nm wide bands used in their PMT spectroradiometry setup: interpolation of voltage versus temperature from a tungsten lamp heated to a wide range of temperatures (Fig. 3 of Ref. \onlinecite{Zhang2015}).

We find the best fit values of~$\epsilon$ and~$T$ by fitting (\ref{eqn:TwoParamPlanck}) with the ``optimize.curve\_fit'' function within the scipy library in python. We estimate relative uncertainties in the fitted data,~$dV/V$, as the inverse square root of the absolute value of measured voltage,~$|V_i|^{-0.5}$, in order to account for shot noise in the photocathode of the MA-PMT assembly. (I.e., we set ``absolute\_sigma = False'' and sigma~$=|V_i|^{-0.5}$ in optimize.curve\_fit).

\subsection{Step 2 of temperature analysis: a series of one-parameter fits}
The second step in our temperature analysis is a one parameter fit to the total measured voltage, assuming the best fit emissivity from step 1.  We numerically solve the following equation for temperature,~$T$, at each time,~$t$:

\begin{equation}
    \epsilon \int{B( T, \lambda) \sum_{i}{M(\lambda,i) d\lambda}} = \delta\lambda \sum_{i}{V_i}
    \label{eqn:one_color_T}
\end{equation}

The numerical method to solve (\ref{eqn:one_color_T}) at each time is simply the interpolation of the function on the left hand side to the computed value of the right hand side.

\begin{figure*}[tbhp]
	\centering
	\includegraphics[width=.95\linewidth]{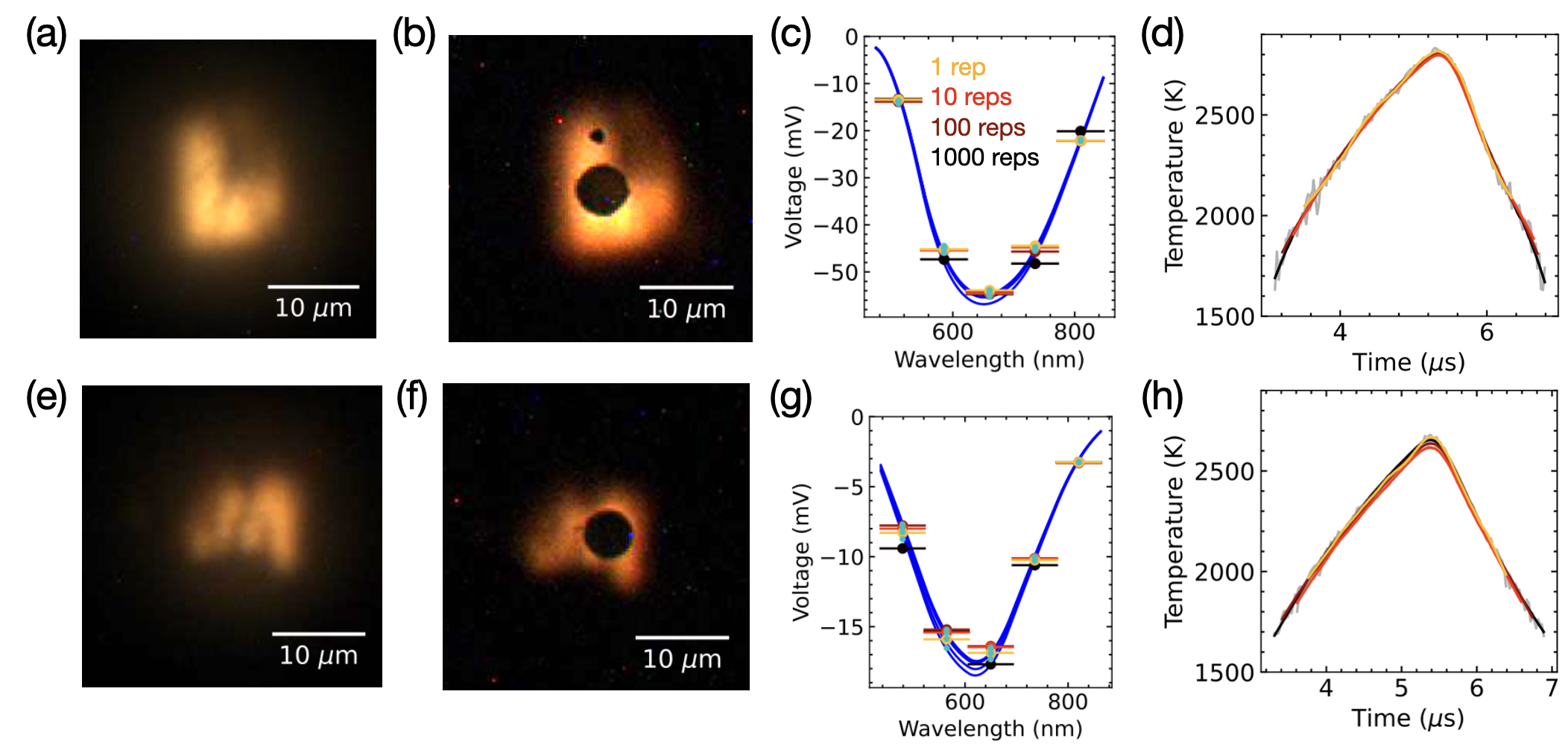}
	\caption{Photographs and temperature measurements for four pulsed-Joule heating experiments of iron at~$\sim 76$~GPa. The top and bottom rows show data from the left and right side of the optical table, respectively. The left and right sides of the table collect thermal emissions from the cylinder-side and piston-side of the DAC. (a, b, e, f) Photographs of the cumulative thermal emissions during 100 heating repetitions, viewed directly in CMOS cameras C1 and C2 (a and e), and after reflection off the mirror pinhole in CMOS cameras C3 and C4 (b and f). (c, g) Wavelength-dependence of MA-PMT voltages from thermal emissions collected during the time period 5-5.5~$\mu$s during a set of 1000 heating repetitions (black circles), 100 heating repetitions (dark red circles), 10 heating repetitions (bright red circles), and 1 heating repetition (yellow circles). Horizontal line segments indicate the width of wavelength bands. Blue curves and cyan dots represent the best fit temperature and emissivity, with blue curves marking the integrand in equation \ref{eqn:TwoParamPlanck},~$M B(\epsilon,T)$ and cyan dots showing the average value of~$M B(\epsilon, T)$~across each band of wavelengths. (d, h) Temperature evolution during each set of heating experiments. The filter timescale is 0.6~$\mu$s.}
	\label{fig:Fe_example}
\end{figure*}

Before plotting, we typically smooth the curves~$T(t)$~using a Savitkzy-Golay filter with polynomial order 2 and timescale,~$\tau$, that is at least 5 times smaller than the pulse duration (e.g.~$\tau = 0.6$~$\mu$s for the 5~$\mu$s pulse in Fig. \ref{fig:Fe_example}). In all cases, we plot the filtered values of~$T(t)$~in colors. In some cases, we plot the unfiltered~$T(t)$~curves in grey (e.g. Fig. \ref{fig:Fe_example}).


\begin{figure*}[tbhp]
    \centering
    \includegraphics[width=4in]{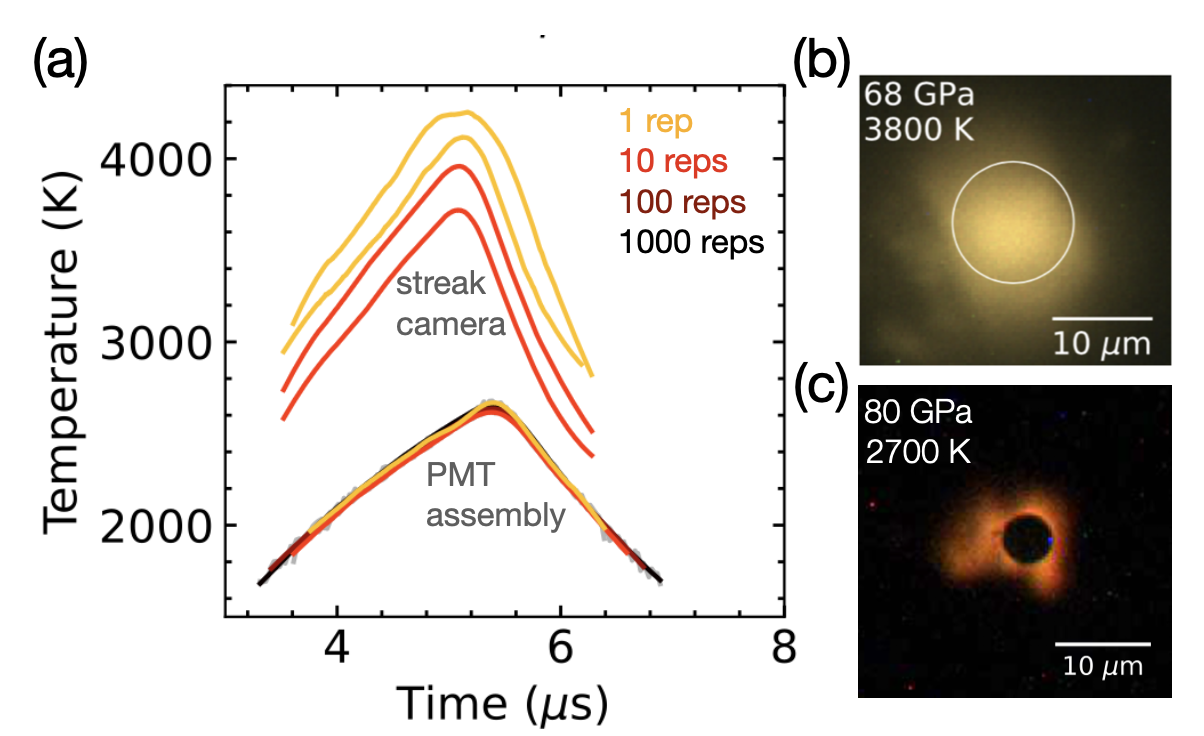}
    \caption{Streak camera versus a MA-PMT assembly. (a) A comparison of temperature evolution of a platinum sample heated to~$\sim 4000$~K, 68 GPa, based on the streak camera measurements in Ref. \onlinecite{Geballe2021} versus the temperature evolution of the~$\sim 80$~GPa iron sample heated to 2800 K using a MA-PMT assembly in this study. The platinum curves are newly processed temperature fits using the same data set used to generate Fig. 6 of Ref. \onlinecite{Geballe2021}. Here, we fit temperatures at a wider range of times -- the maximum time range possible without fitting noise and generating~$\sim 1000$~K scatter. The iron curves are identical to those in Fig. \ref{fig:Fe_example} of this study. In both cases, the filtering timescale is 0.6~$\mu$s. (b) Image of the hot platinum sample used to generate the streak camera curves in (a). The white circle marks the region of sample whose emissions are collected on the streak camera. (c) Image of the hot iron sample used to generate the ``MA-PMT assembly'' curves in (a), reflected off a mirror pinhole.}
    \label{fig:streak-v-PMT}
\end{figure*}

\section{Result from a high pressure Joule-heated sample}
To demonstrate the utility of the optical table and MA-PMT detectors for high pressure experiments, we show examples of temperature measurements during pulsed Joule heating of an iron sample compressed in a DAC to~$\sim 70$~GPa at 300 K and heated to~$\sim 2800$~K. Joule heating is performed by the method described in Ref. \onlinecite{Geballe2023Pulser}. Pressure is measured by X-ray diffraction and by the diamond anvil Raman edge.\cite{Akahama2006}

Results of the temperature analysis for data accumulated during four different experiments are shown in Fig. \ref{fig:Fe_example}. In one experiment, a single heating pulse is recorded. In the others, thermal emissions from repeated pulses are accumulated 10, 100, or 1000 times. In all cases, the~$T(t)$~curves show outstanding reproducibility -- the peak temperature varies by less than~$\pm 30$~K on each side of the sample. The~$T(t)$~curves show similarly small scatter down to low temperatures, reaching~$\sim 2000$~K for the single shot data and~$\sim 1700$~K for the 1000 repetition data.

The outstanding signal to noise at temperatures below 3000 K suggests a major improvement compared to our previous study using a streak camera,\cite{Geballe2021} a study that itself pushed the limits of low temperature ($T\sim 3000$~K) for DAC heating experiments with high time resolution ($\sim 0.6$~$\mu$s) and few repetitions (1 to 10). Fig. \ref{fig:streak-v-PMT} compares examples from Fig. \ref{fig:Fe_example}d with one of the most precise and reproducible data sets from Ref. \onlinecite{Geballe2021}, the platinum melting data at~$\sim 68$~GPa. The comparison shows that the precision and shot-to-shot reproducibility documented here at~$\sim 2000$~K is approximately equal to that documented in Ref. \onlinecite{Geballe2021} at~$\sim 3000$~K. 

\begin{figure*}[tbhp]
	\centering
	\includegraphics[width=5in]{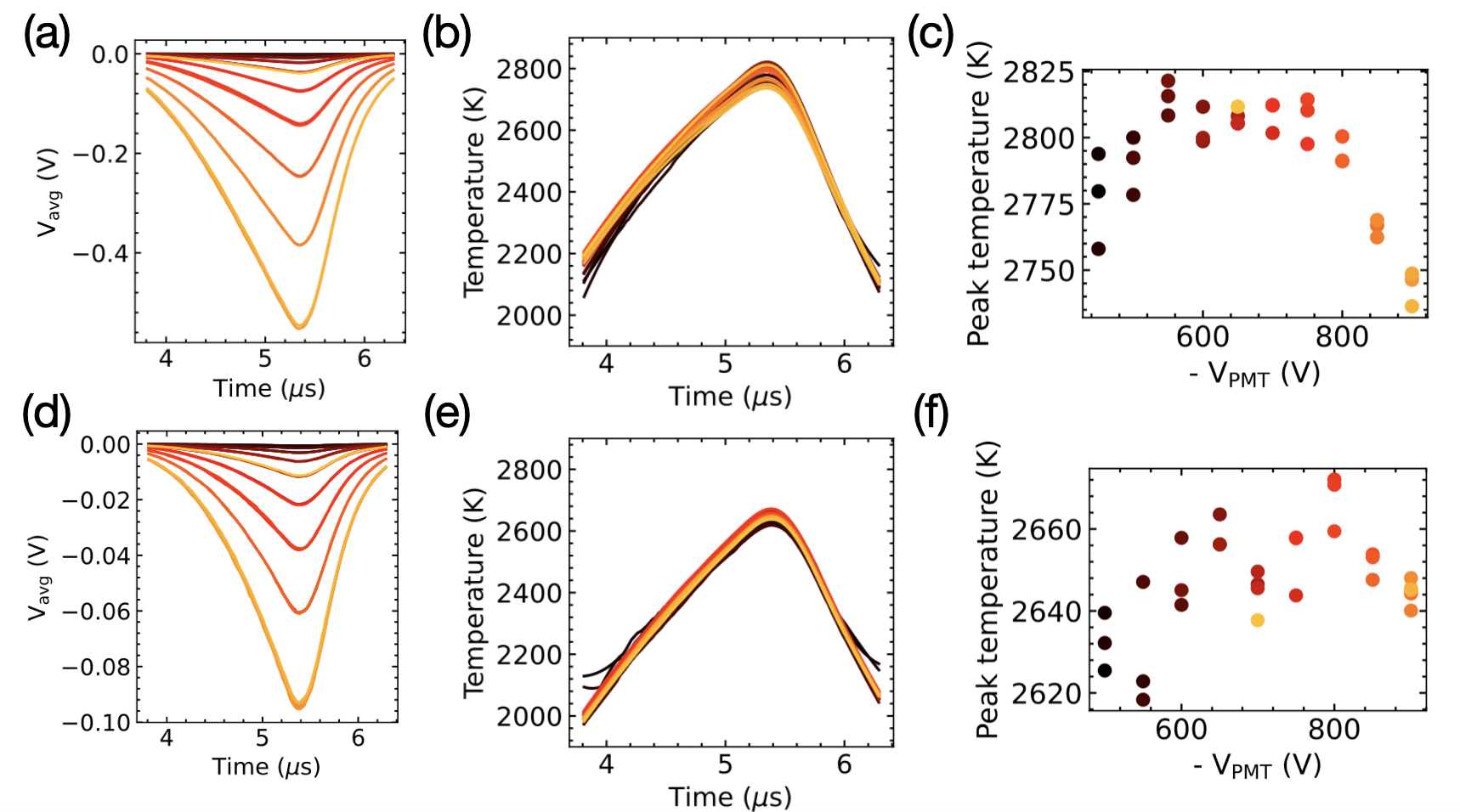}
	\caption{Gain-dependence of temperature fits. Temperature measurements on the left side (top row) and right side (bottom row) of the optical table during pulsed heating of the same high pressure iron sample documented in Fig. \ref{fig:Fe_example} but with a higher heating power. (a,d) Average MA-PMT voltages on each side during 31 heating experiments using variable MA-PMT gain. (b,e) Fitted temperatures versus time. The first 30 experiments were conducted with increasing supply voltage from -450 V to -900 V (black to yellow) yielding a~$\sim100$-fold variation in MA-PMT gain. The 31st experiment (yellow) was collected at -650 V to test for any irreversible changes in the sample. (c,f) Peak temperature versus MA-PMT supply voltage.}
	\label{fig:Fe_gain_scan}
\end{figure*}

Since non-linearity in a PMT response would appear as an insidious smooth function (as opposed an easy-to-detect flat distribution of saturated counts on a CCD), it is important to verify that the thermal emissions data is collected using a gain where the MA-PMT response is sufficiently linear to avoid systematic error larger than the claimed precision ($\pm 30$~K). Indeed, temperature measurements on high pressure iron are independent of the gain of the MA-PMT assembly within a 30-fold range of gain, for both left and right side. On the right side, there is no detectable gain-dependence for supply voltages above -600 V, within the~$\pm 20$~K scatter (Fig. \ref{fig:Fe_gain_scan}f). On the left side, there is gain-dependence to peak temperature within the range -550 to -800 V (Fig. \ref{fig:Fe_gain_scan}c), corresponding to averaged PMT voltages in the range -8 to -200 mV, a 30-fold range. Above -800 V, the non-linear response of the PMT at high anode current degrades accuracy, causing anomalously low temperature fits by up to 60 K at -900 V. Below -550 V, the thermal emissions data becomes very noisy.

\begin{figure*}[tbhp]
    \centering
    \includegraphics[width=5in]{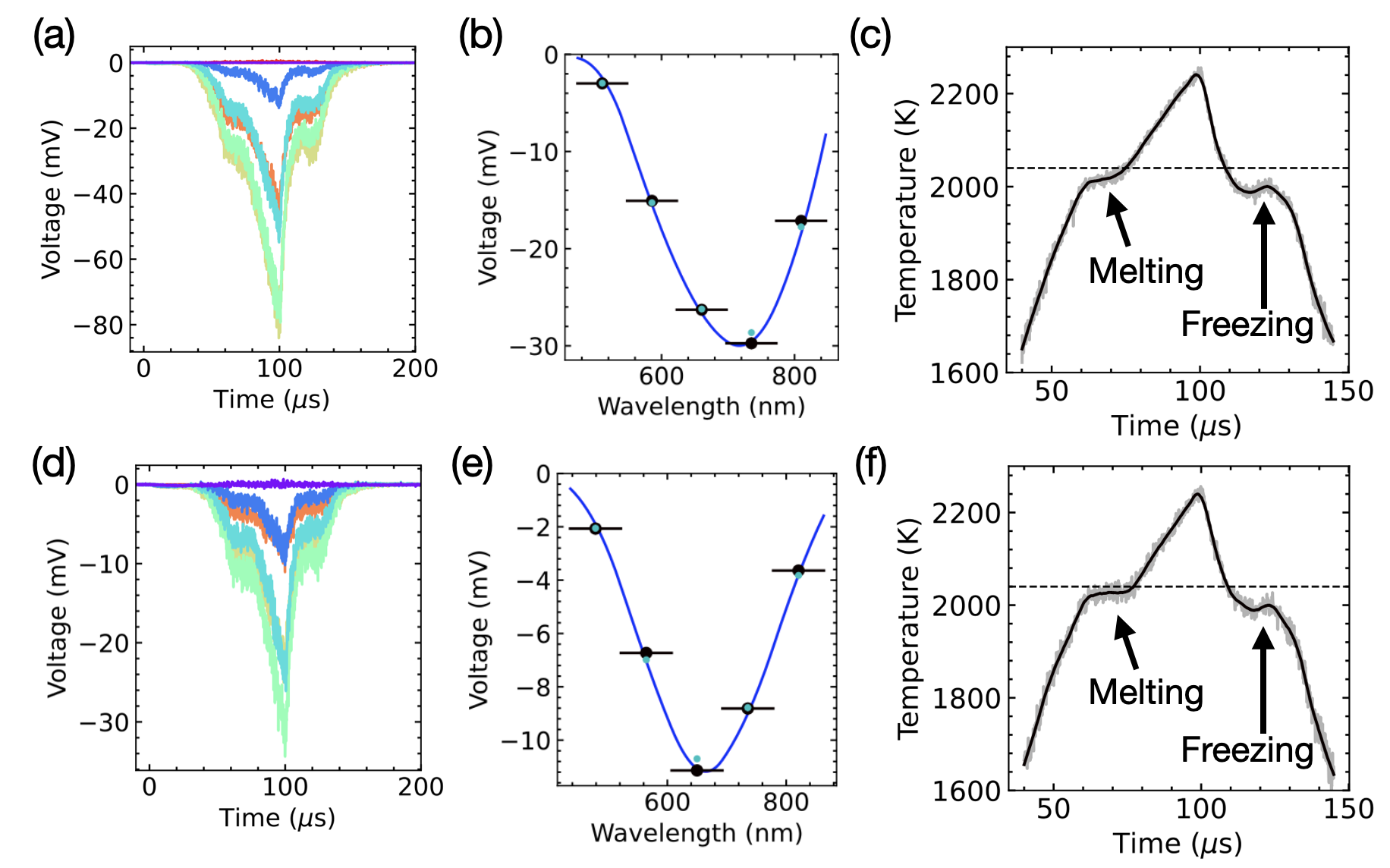}
    \caption{Melting and freezing of a platinum sample at~$P \approx 1$~atmosphere. The sample is a wire of platinum (Alfa Aesar 10292, 25~$\mu$m diameter, 99.95\% Pt) that is pressed between~$\sim50$~$\mu$m-thick layers of KCl which are pressed gently between diamond anvils (i.e. a DAC with no gasket, compressed to less than 1 GPa of pressure). (a,d) Oscilloscope traces after subtracting the spurious electrical pickup. The rainbow of colors indicates the spectral bands measured in the oscilloscope, matching Fig. \ref{fig:wavelength_calibration}, from orange at~$\sim 800$~nm to blue at~$\sim 550$~nm. (b,e) Circles show the measured voltage spectrum averaged over the time range 67-75~$\mu$s. As in Fig. \ref{fig:Fe_example}, horizontal line segments indicate the width of wavelength bands. The cyan dots and blue curves show the values of voltage for the best fit temperature and emissivity. Cyan dots show the average across each wavelength band, while blue curves show the continuous spectrum of voltages,~$M B(\epsilon,T)$. Fit parameters and uncertainties on left (b) and right (e) are~$T = 2025 \pm 31$~K,~$\epsilon = 0.42 \pm 0.07$, and~$T = 2029 \pm 32$~K,~$\epsilon = 0.29 \pm 0.05$. (c,f) Temperature versus time on the left and right sides. Arrows point to plateaus cause by the latent heat of fusion upon melting and freezing. Ten repetitions are averaged. The filter timescale is 10~$\mu$s.}
    \label{fig:Pt_example}
\end{figure*}

\section{Ambient pressure melting tests}
We use ambient pressure melting experiments to test the accuracy of spectroradiometric temperature measurements using the new optical table and MA PMT-based detectors. We report tests of four materials that melt in the range 1808 -- 2719  K: iron, platinum, alumina, and iridium. 

In all cases, melting is detected by unambiguous plateaus upon heating to temperatures at or above the melting temperature. An example is shown in Fig. \ref{fig:Pt_example} where several platinum melting experiments are documented. Upon heating, the temperature evolution on each side is nearly linear before plateauing during the time interval 67-75~$\mu$s. Each side plateaus at a temperature within 20 K of the known melting temperature of platinum, 2040 K. 

After the melting plateau, the temperature on each side increases by~$\sim 200$ K, and decreases sharply after the current pulse ends. Upon cooling, the temperature on each side again plateaus at approximately the same temperature as the heating plateau. The heating plateaus are caused by absorption of the latent heat of fusion; the cooling plateaus are caused by the release of the latent heat of fusion. Since supercooling of fluids is a much more common phenomenon than superheating of solids,\cite{Ubbelohde1978} we focus on the melting plateaus. For example, in step 1 of the temperature analysis, we use the melting plateau at 67-75~$\mu$s as the time-interval of interest.

While it is tempting to also interpret the best fit values of emissivity at melting, in most cases, the fitted value of emissivity includes an error of order 10s of percent caused by the system response calibration. We typically calibrate the system response at a different gain than the measurement -- a difference that is often required because of the differing ranges of linearity in pulsed versus continuous PMT measurements. We correct the system response for an approximate gain function that leaves a residual uncorrected gain of order 10\%:~$V_\textrm{W,corrected} = V_\textrm{W} \times 10^{-\Delta V_\PMT/140 \textrm{~V}}$ where~$\Delta V_\PMT$ is the difference in MA-PMT supply voltage between tungsten lamp and sample measurements.

\begin{figure*}[tbhp]
    \centering
    \includegraphics[width=5in]{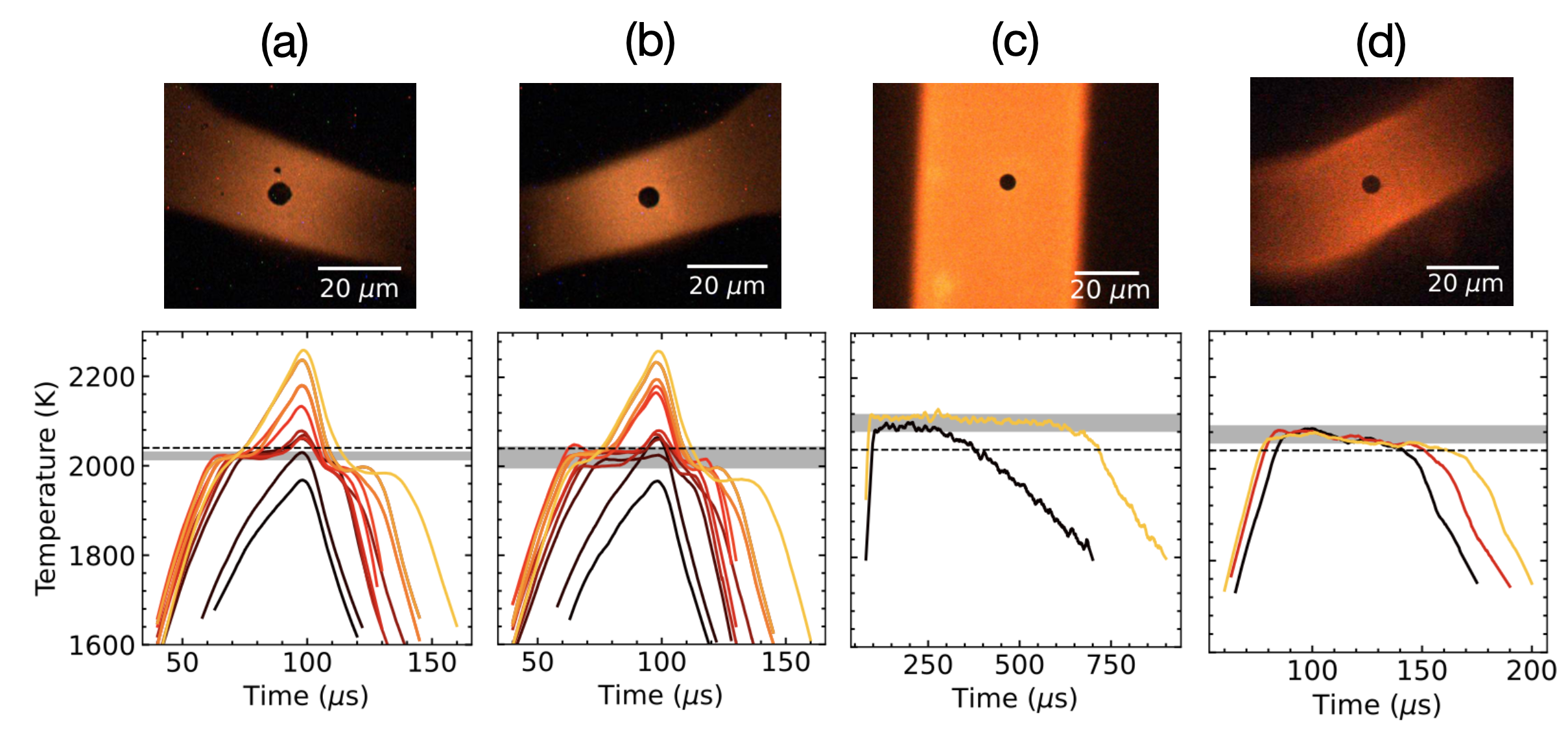}
    \caption{Melting and freezing of platinum samples at~$P \approx 1$~atmosphere. (Top) Images of hot samples on mirror pinholes during individual melting experiments. The black circle in the middle of each image is a mirror pinhole. The small black spot above the mirror pinhole in (a) is an imperfection in the mirror. (Bottom) Temperature evolution during multiple heating experiments on each sample. Each color from black to yellow represents one set of heating experiments. Dashed lines mark the literature value of platinum's ambient pressure melting temperature, 2040 K. Grey shading marks our estimate of plateau temperature. (a,b) Data from the left and right sides of the same sample shown in Fig. \ref{fig:Pt_example} -- platinum trapped between layers of KCl, 10 repetitions per measurement, filtering timescale 10~$\mu$s. (c) Data from a piece of 50~$\mu$m-diameter platinum wire heated in vacuum, using a single pulse of current (i.e. 1 repetition). The filtering timescale is 20~$\mu$s. (d) Data from a piece of 25~$\mu$m wire heated in air, using a single pulse of current. The filtering timescale is 15~$\mu$s.}
    \label{fig:Pt_many}
\end{figure*}

The same platinum sample from Fig. \ref{fig:Pt_example} was melted repeatedly. Fig. \ref{fig:Pt_many}a,b documents repetitions of the melting measurements. The range of melting plateaus is~$2022 \pm 10$~K and~$2020 \pm 25$~K on the left and right sides of the optical table (shaded grey regions). Figs. \ref{fig:Pt_many}(d,e) show two other platinum samples that were melted at ambient pressure. Those two samples were pieces of 50 or 25~$\mu$m-diameter wire pressed into thinner copper electrodes and melted in air (i.e., we did not use a KCl medium or a DAC to hold the sample). The discrepancy with respect to the 2040 K melting temperature of platinum are slightly greater for these samples (+35 K and +60 K, rather than -20 K for the sample in KCl).

\begin{figure}[tbhp]
    \centering
    \includegraphics[width=2.5in]{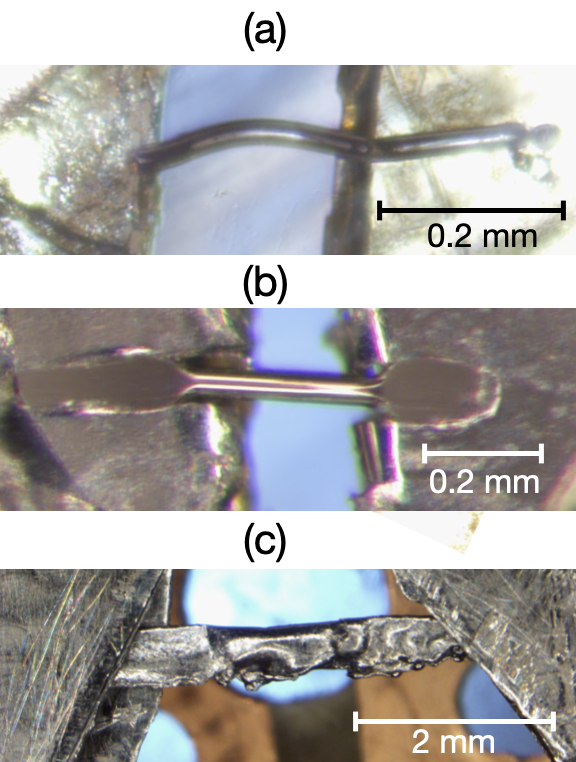}
    \caption{Examples of ambient pressure samples. (a) A platinum sample fixed between anvils with 1 mm culets, no gasket, and KCl insulation. Corresponding data is in Fig. \ref{fig:Pt_example}. (b) A platinum sample pressed onto a copper clad board that is coated in a thin layer of solder. Anvils were used to press the platinum into the solder, and then removed, leaving the platinum embedded in the solder. Corresponding data is in Fig. \ref{fig:Pt_many}c. (c) The rhenium capsule filled with alumina, welded shut, and clamped between pieces of molybdenum foil. Corresponding data in Fig. \ref{fig:Al2O3_many}d,e.}
    \label{fig:ambient_gallery}
\end{figure}

For ambient pressure melting experiments on metals in air or in vacuum, a wire of the desired metal was pressed into a copper clad board previously divided into a negative and positive side by cutting a narrow trench with a razor blade. The pressing action was created by using our hands to firmly clamp down on a diamond cell equipped with anvils with 1 mm culets (or in some cases, 0.5 mm culets). Pressing for a couple seconds and then releasing was sufficient to fix the metal in place for iridium. For the relatively soft platinum and iron samples, a thin solder coating was added on top of the copper before pressing the sample. No adhesive was used to fix the wires. Examples of the ambient pressure samples are in Fig. \ref{fig:ambient_gallery}.

\begin{figure*}[tbhp]
    \centering
    \includegraphics[width=5in]{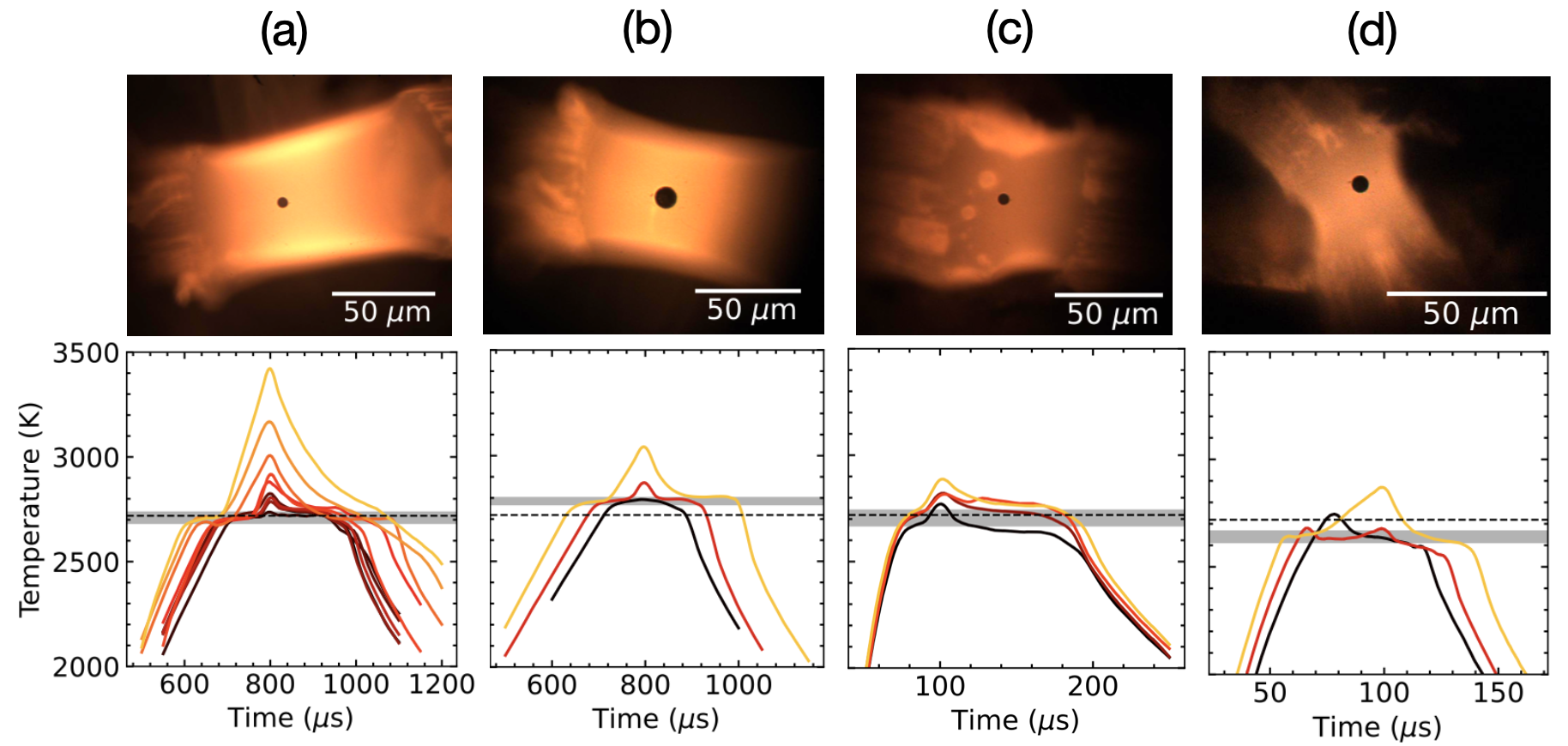}
    \caption{Melting and freezing of iridium samples at~$P \approx 1$~atmosphere. (Top) Images of the hot sample on the mirror pinhole during an individual melting experiment. (Bottom) Temperature evolution during multiple heating experiments on each sample. Each heating experiment uses a single pulse. (a,b) The same iridium sample melted in air using (a) the standard 200 mm focal length focusing lens, and (b) a 100 mm focal length focusing lens that lowers the system magnification 2-fold. The outer edges of the images show the texture of un-melted iridium, whereas the central~$\sim 100$~$\mu$m wide region shows the texture of iridium after melting and refreezing multiple times. (c) A different iridium sample melted in air. The bright spots in the image formed during repeated melting experiments. (d) An iridium sample melted under vacuum.  The filtering timescale for (a)-(d) are 40~$\mu$s, 50~$\mu$s, 20~$\mu$s, and 10~$\mu$s.}
    \label{fig:Ir_many}
\end{figure*}

\begin{figure}[tbhp]
    \centering
    \includegraphics[width=3in]{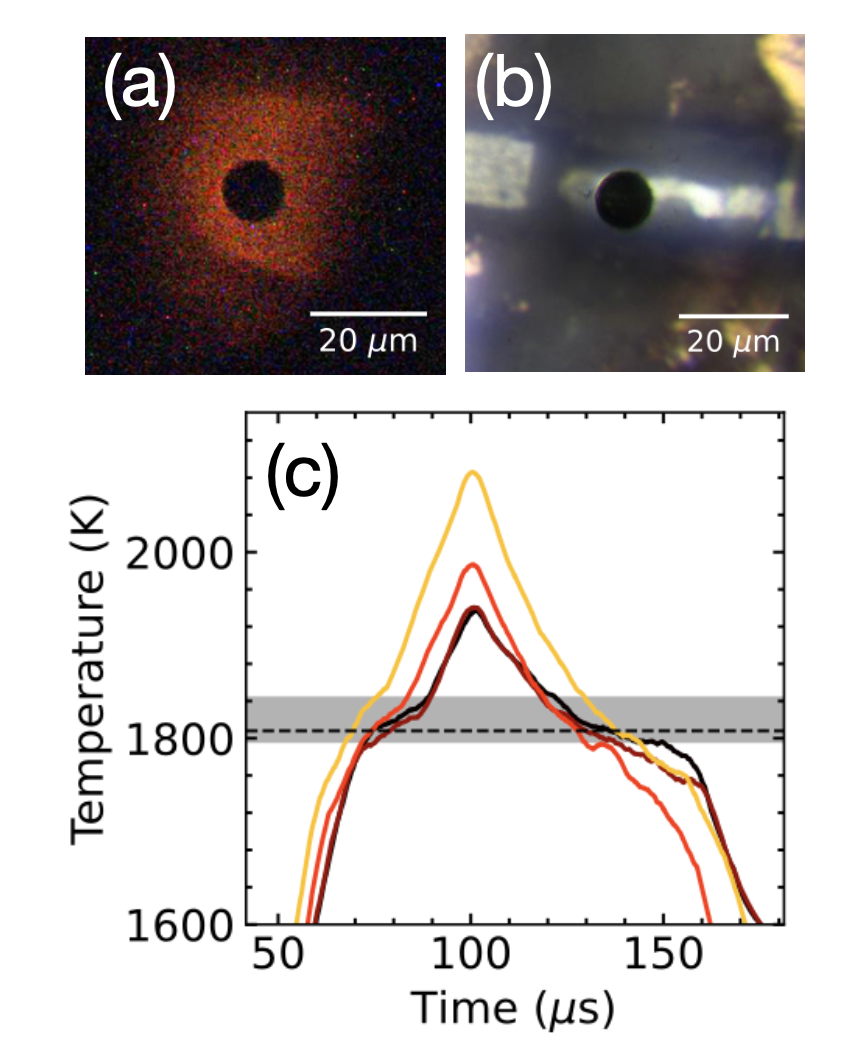}
    \caption{Melting and freezing of iron under vacuum. Images of the mirror pinhole during melting (a) and in white light after melting (b). (c) Temperature evolution during multiple heating experiments. Each experiment uses a single pulse. The filtering timescale is 10~$\mu$s.}
    \label{fig:Fe_many}
\end{figure}

\begin{figure*}[tbhp]
    \centering
    \includegraphics[width=6in]{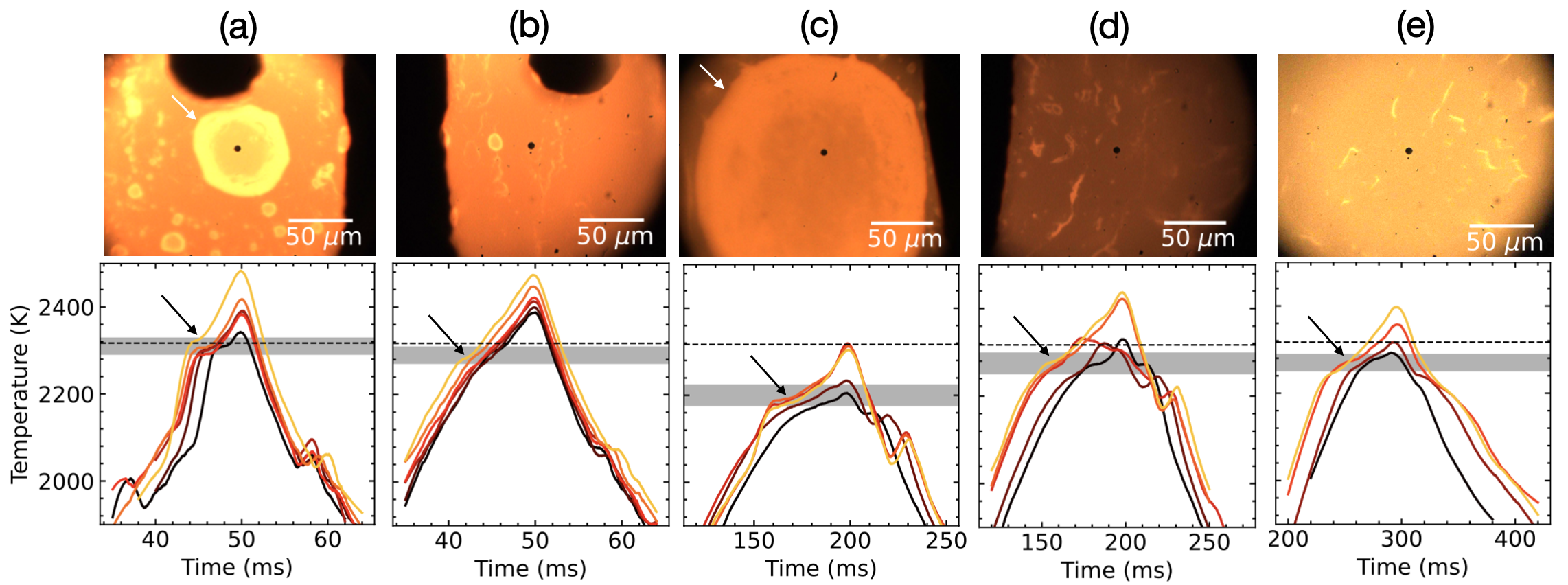}
    \caption{Melting and freezing of alumina heated by conduction from Joule heated rhenium under vacuum. (Top) Images of the rhenium and alumina on mirror pinholes. White arrows in (a) and (c) point to the edge of alumina blobs. (Bottom) Evolution of color temperature, a proxy for the rhenium temperature. Black arrows point to plateaus caused by melting of alumina. (a,b) A flattened rhenium wire with a disc of alumina on one side of the wire (yellow disc in image (a)). (c,d) A different rhenium wire, flattened, with a pile of alumina on the side in (c). (e) The surface of an rhenium capsule stuffed with alumina. Each experiment uses a single pulse. The filtering timescales for (a)-(e) are 3 ms, 3 ms, 15 ms, 15 ms, and 30 ms.}
    \label{fig:Al2O3_many}
\end{figure*}

The data for melting of iridium, iron, and alumina are shown in Figs. \ref{fig:Ir_many}, \ref{fig:Fe_many}, \ref{fig:Al2O3_many}. The iron and one of the iridium samples were mounted on a copper clad board and inserted into a vacuum chamber with glass viewports. (System response calibration with and without the glass viewport showed no measurable difference in spectral shape). The iridium samples were cleaved from a piece of iridium wire (Alfa Aesar 11430, 99.8\% Ir, 0.5 mm diameter). The iron sample was a piece of 25~$\mu$m diameter iron wire (Goodfellow 030-530, 99.99\% Fe).

The alumina samples were placed inside or on top of a rhenium heater, and held inside the vacuum chamber during heating. In the first experiment, a rhenium heater was constructed from a flattened piece of 100~$\mu$m-diameter rhenium wire (Alfa Aesar 10310-BU, 99.97\%), which was clamped between pieces of brass. The brass clamp was made by soldering brass to a copper clad board. A disc of alumina was created by the following serendipitous procedure: a 100~$\mu$m hole was drilled in the rhenium and filled with pieces of alumina~(Johnson Matthey, 22~$\mu$m mesh), which were then melted, creating a disc of alumina. The disc of alumina jumped out of the hole and landed on the wire in the location shown in Fig. \ref{fig:Al2O3_many}b. The thermal emissions from the alumina-covered rhenium show a jump in PMT signal when the alumina melts and becomes transparent, allowing increased transmission from the Re. After the jump, the temperature evolution is anomalously for a few ms, before rising again. This evolution suggests that the apparent plateau in temperature is caused by the latent heat of melting. Note that it is implausible for the plateau to be caused by a change in the rhenium heater, which is expected to melt 3455 K, and which does not undergo any textural changes visible in white light images. The other side of the rhenium heater also shows a subtle plateau at the same time, a result of the latent heat of melting but with a substantial temperature difference between the other side's surface and the alumina (Fig. \ref{fig:Al2O3_many}a). Both sides show plateaus due to the latent heat of freezing with large supercooling, evidenced by 200 K of hysteresis.

Our attempt to reproduce this serendipitous sample preparation, resulted in a much larger pool of alumina on one side of the new rhenium heater (Fig. \ref{fig:Al2O3_many}c). The alumina side shows a less reproducible jump in PMT signal upon melting, and much less transparent alumina in the heating photos, perhaps because part of the alumina blob did not melt. Still, the alumina side shows a plateau in temperature at~$2200 \pm 25$ K, far below the 2317 K melting temperature of alumina. The discrepancy is likely because emissions from the rhenium undergo wavelength-dependent scattering by grain boundaries in the crystalline regions of the alumina. On the other hand, the clean side of the rhenium heater shows a subtle melting plateau at temperature slightly lower than than 2317 K (Fig. \ref{fig:Al2O3_many}c), and a sharp freezing plateau with~$\sim 50$~K of hysteresis. 

For the final experiment on alumina, we used common equipment for multi-anvil experiments to make a rhenium capsule with 50~$\mu$m-thick rhenium foil, which we stuffed with alumina pieces and closed by crimping and spark welding. We clamped the edges of the rhenium capsule between pieces of 0.5 mm thick Mo foil. The result is an ambient pressure heater and sample with larger mass than the other samples documented here, requiring 100s of ms to heat. It shows clear plateaus upon melting and freezing at temperature slightly lower than the known value of melting temperature of alumina (Fig \ref{fig:Al2O3_many}e). The slightly depressed plateau temperatures for rhenium surfaces that are not touching alumina (Fig. \ref{fig:Al2O3_many}a-d) are likely caused by temperature gradients between the location on the rhenium that is measured and the molten alumina. Indeed, we document no discrepancy in the one example in which the measured rhenium surface is covered in a disc of fully molten alumina.

Finally, we mention several failed attempts to document accurate melting temperatures using the new MA-PMT system. First, heating tungsten or tantalum in air resulted in plateau temperatures that ranged from 100 to 800 K below the literature values of melting temperature. Second, after nearly 100 repetitions of melting the iridium shown in Fig. \ref{fig:Ir_many}a,b in air, the color temperature of the plateau decreased by $\sim 100$ K to 2620 K. Third, heating rhenium and tungsten in KCl resulted in depression of plateau temperatures by 300-400 K. In all cases, chemical reaction with KCl or with water absorbed in the KCl could cause the discrepancy by lowering the melting temperature, or by introducing a substantial wavelength dependence of emissivity. 


\begin{figure}[tbhp]
    \centering
    \includegraphics[width=3.3in]{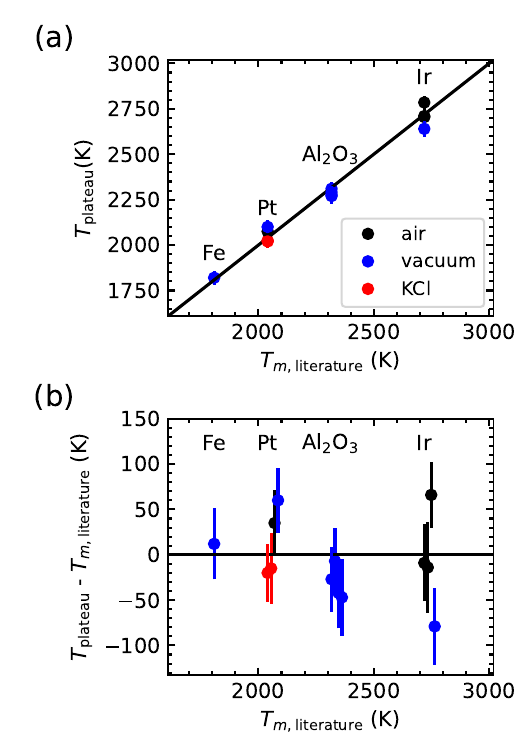}
    \caption{Summary of the precision and accuracy of temperature measurements at ambient pressure. (a) Measured plateau temperature versus literature value of melting temperature for iron, platinum, alumina, and iridium. Colors indicate the medium in which the sample is heated: air (black), vacuum (blue), or KCl (red). (b) The same data in (a) after subtracting~$T_{m,\textrm{literature}}$. Each symbol and error bar represent the average plateau temperatures and the quadrature sum of two sources of uncertainty: the variation in plateau temperature (i.e. the half-width of the grey shading in Figs. \ref{fig:Pt_many}-\ref{fig:Al2O3_many}), and the 30 K uncertainty ($1\sigma$) that is typical of two parameter Planck fits in this study.}
    \label{fig:accuracy_summary}
\end{figure}

Fig. \ref{fig:accuracy_summary} summarizes the fifty-five plateaus documented on ten distinct samples in Figs. \ref{fig:Pt_example}-\ref{fig:Al2O3_many}. (We omit the anomalously low plateau temperatures measured when viewing through a large blob of alumina in Fig. \ref{fig:Al2O3_many}c.) Each set of melting experiments (e.g. Fig. \ref{fig:Pt_many}a) is represented by a single data point in Fig. \ref{fig:accuracy_summary}. In all cases, the average plateau temperature measured for a sample is within 80 K of the literature value for melting temperature. This suggests that the accuracy of the temperature measurement is better than~$\pm80$~K. The rms deviation is 40 K, suggesting a 1$\sigma$~accuracy of 40 K.

\section{Discussion}

The typical precision of the MA-PMT based spectroradiometry system is~$\pm 30$~K. The accuracy may be as good as~$\pm 40$~K (1$\sigma$), but a more conservative estimate is the~$\pm 80$~K range that encompasses all plateaus documented here. The rise time is 0.24~$\mu$s, but we typically filter to 0.6~$\mu$s for the high pressure data and to 10s or 100s of~$\mu$s for the slower heating to lower peak temperatures in our ambient pressure melting tests. 

The system can be calibrated using common optical tools, plus the tungsten lamp that is a necessary tool for spectroradiometry. The calibrations and temperature fitting procedures are more complicated than the procedures for standard spectroradiometry because of the mediocre spectral resolution of the MA-PMT assembly. An alternative design with~$\sim 4$~times higher spatial resolution is possible. It would require four times as many digitizer channels, and it would allow for a simpler data analysis without the interpolation step. 

The main result of this manuscript is that MA-PMT assembly can be integrated into a time-resolved spectroradiometry system for DAC experiments. Moreover, the data suggest that MA-PMT assembly provides a major improvement in signal-to-noise compared to streak cameras. Our implementation of a MA-PMT assembly provides a substantial improvement in signal to noise over our previous work using a streak camera detector,\cite{Geballe2021} at least at low signal intensity. For comparison, Fig. \ref{fig:streak-v-PMT} shows that our the MA-PMT-based optical table is capable of measuring precise temperatures down to~$\sim 1900$~K and 2000 K when accumulating 1 and 10 repetitions (lower red and yellow curves), whereas the streak camera-based optical table used in Ref.~\onlinecite{Geballe2021} is capable of measuring precise temperatures down to~$\sim 2600$~and 3000 K (upper red and yellow curves). The difference in temperature corresponds to a factor of 20 in number of photons measured per second (assuming the system response of the MA-PMT-based optical table), suggesting that the MA-PMT-based system is approximately 20 times more sensitive than the streak camera based system. Moreover, our MA-PMT-based optical system is designed to have lower chromatic aberration (4.5 mm vs. 8 mm aperture diameter, using the same 20x NIR Mitutoyo objective) and higher resolution spatial filtering (5~$\mu$m~vs. 12~$\mu$m diameter collection area). These two differences in design cause the new system's spectrometers to collect~$3.26 \times \left(\frac{12}{5}\right)^2 = 18$~times less light than the system in Ref.~\onlinecite{Geballe2021}, assuming a light source that is uniform over a 12~$\mu$m area. (The factor of 3.26 is the difference in solid angle subtended by the full versus apertured objective lenses). All together, this suggests that 5-color spectroradiometry based on a MA-PMT assembly is~$18 \times 20 = 360$-times more sensitive than the approximately 800-color spectroradiometry based on streak cameras in Ref.~\onlinecite{Geballe2021}. 

The 360-fold increase in sensitivity opens up major opportunities to study Earth's mantle and core during pulsed heating experiments. For example, thermal conductivity measurements at lowermost mantle conditions may be possible in single-shot mode, eliminating the need for reproducible heating during 100s to 1000s of repetitions (as used in Refs.~\onlinecite{Konopkova2016,McWilliams2015,Geballe2020}). Likewise, the improved sensitivity expands the measurable temperature range during single-shot pulsed heating DAC experiments on chemically reactive samples such as fluid H$_2$O or H$_2$.

There are two likely causes for the 360-fold increase in sensitivity. First, the much lower number of colors measured in the MA-PMT system reduces noise since each channel of a detector's readout electronics adds its own dark noise. Second, individual PMT channels may have higher quantum efficiency or lower noise than the phosphor detector and multichannel plate that make up the front end of the streak camera. Regardless of the reason, MA-PMT assemblies provide a major improvement in signal-to-noise over the streak camera used in Ref. \onlinecite{Geballe2021}, at least at the low thermal emission intensities achieved with a 2000-3000 K source, 1 to 10 repetitions, and a 0.6~$\mu$s filtering timescale.

Since streak cameras themselves are not a widespread technology in DAC labs, we briefly compare MA-PMTs and streak cameras with other technologies that are used for spectroradiometry in DACs: iCCDs and sets of single-channel photodiodes or PMTs. The major advantage of a MA-PMT or streak camera over an iCCD is that thermal emissions are measured during a \textit{range of} times during each pulse, rather than averaging over a single time interval. This means that a measurement of temperature at 20 different times requires 20-times more repetitions than needed in a MA-PMT or streak camera system. This is a crucial difference for samples that do not heat and cool reproducibly (e.g. because of shape changes during melting). Moreover, MA-PMTs and streak cameras have two main advantages compared to the sets of single channel detectors used in Refs. \onlinecite{Radousky1989,Zhang2015,Montgomery2018,Li2020,Bassett2016,Boslough1989}. First, alignment of different wavelengths is trivial since a spectrometer can focuses all wavelengths at the location of the PMT anodes. Second, high throughput of light is easy to achieve because of the high efficiency of the spectrometer's grating. In other words, a MA-PMT assembly or streak camera couples well with a spectrometer, enabling the advantages of spectrometers in spectroradiometry. One limitation of MA-PMT assemblies is that few options are available in the market. For example, we are not aware of any MA-PMT assemblies with sensitivity in the mid-IR spectrum, unlike the single channel detectors used in Ref \onlinecite{Montgomery2018}. Another disadvantage of the designs using an MA-PMT assembly plus spectrometer rather than a set of single channel detectors and bandpass filters is that wavelength calibration is trickier. On the other hand, the near linearity of dispersion of modern spectrometers enables the simple calibration routine employed here. In practice, the MA-PMT assembly based system presented here has a much higher time resolution (0.24~$\mu$s rise time) than the single channel detectors used in Ref. \onlinecite{Montgomery2018} (14~$\mu$s sampling rate; 5 kHz InGaAs detectors) and Ref. \onlinecite{Zhang2015} (20 kHz working frequency), though not as high as streak cameras, which can achieve ns time resolution. 

\section{Conclusions}
A new optical system for spectroradiometry in pulsed Joule heated diamond cells is presented. MA-PMT assemblies provide outstanding precision and accuracy for spectroradiometry at~$\sim 1700$~to 2700 K and sub-$\mu$s time resolution while collecting from a 5~$\mu$m-diameter measurement area. The system is compatible with Joule heated DACs. The precision and accuracy of the new system are estimated to be ~$\pm 30$~K and~$\pm 80$~K.

\begin{acknowledgments}
This material is based upon work supported by the National Science Foundation under Grant No. 2125954. We thank Seth Wagner, Vic Lugo, and Cesar Sanchez for machining parts. We thank Amol Karandikar and Joe Lai for fruitful discussions.
\end{acknowledgments}

\section*{Data Availability Statement}
The data that support the findings of this study are available from the corresponding author upon reasonable request. 

\appendix

\section{Mirror pinhole fabrication}
\label{SM:mirror_pinhole}
The mirror pinholes were fabricated from broadband-coated glass discs (Edmund Optics 45-658). First a piece of platinum wire was pressed to 5~$\mu$m thickness and laser cut into an $\sim 100$~$\mu$m-diameter disc. Next, the disc was placed at the center of the glass disc. Next, a~$\sim 200$~nm-thick layer of aluminum was deposited on the masked optical glass using RF magnetron sputtering. Next, the platinum disc was removed using a micromanipulator. Next, dust was blown off the glass’s surface using compressed air. This revealed dozens of~$\sim 10-50$~$\mu$m wide holes in the aluminum, spaced randomly around the 12.5 mm-diameter piece of glass. Next, the platinum disc was placed in the same location with~$\sim 2$~$\mu$m accuracy using a micromanipulator. Next, another~$\sim 200$~nm thick layer of aluminum was deposited on the re-masked glass disc. After removing the platinum disc and blowing with compressed air, no holes were observed in the deposition, leaving the intentional pinhole as the only hole. However, imperfections are evident in the aluminum coating, such as the small dot above the pinhole in Fig. \ref{fig:Fe_example}b. The imperfections are the locations that were accidentally masked by dust particles during one of the two depositions, meaning the aluminum is~$\sim 200$~nm thick in those areas.

\section{Linearity of the MA-PMT assembly}
\label{SM:PMT_linearity}
We use the 32 channel MA-PMT assemblies from Hamamatsu because the 32-channel version seems to have much larger dynamic range than the 8-channel version. Our testing with 10~$\mu$s duration pulses of LED light suggests that the 8-channel version saturates at~$\sim 30$~$\mu A$ anode current regardless of supply voltage, whereas the 32-channel version saturates at 50 to 250~$\mu$A for supply voltage from -500 to -800 V, with saturation current increasing with the magnitude of the supply voltage as expected when the mechanism of saturation is that photoelectron current begins to have a large effect on the voltage of the voltage divider. The~$\sim 10$-fold difference in anode saturation current at high gains effectively gives the 32-channel MA-PMT assembly a 10-fold increase in dynamic range compared to the 8-channel version.

The MA-PMT assembly is placed at the spectrometer focus with mediocre accuracy -- ~$\sim 3$~mm accuracy on the Acton spectrometer side (focal length 300 mm), and~$\sim10$ mm accuracy on the Holospec side (85 mm focal length). Whatever misalignment of the MA-PMT assembly position exists in our system does not noticeably degrade our spectral resolution; shifting the focal position of the detector by 10 mm does not change the sharpness of peaks in our wavelength calibration (Fig. \ref{fig:wavelength_calibration}). Moreover, the slight misalignment might be beneficial for spreading light over the MA-PMT’s photodiodes, because according to Ref. \cite{Bassett2016}, this enhances the range in which the PMT channels have a linear response (Fig. \ref{fig:linearity_test}. In an attempt to enhance the dynamic range of the MA-PMT assembly on the Acton spectrometer side, we tested the effect of adding an anisotropic holographic diffuser (Edmund Optics 47-999) located~$5\pm 2$~mm from the PMT photocathodes. The diffuser's high scattering direction was aligned along the height of PMT photocathodes so that at the spectrometer's focal position, the rainbow of light from the 100~$\mu$m diameter pinhole would no longer be focused to its nominal height ($\sim 100$~$\mu$m), but would instead be defocused to~$\sim 3$~mm in height. The measured effect on PMT linearity was negligible, so we removed the diffuser. 


\section{Electrical pickup from Joule heating pulses} 
\label{SM:electrical_pickup}
A large pulse of Joule heating current contributes a spurious component to the voltage measured on PMT readout electronics. For example, a 10 A, 100~$\mu$s heating pulse causes a time varying voltage signal of~$\sim 1$~mV amplitude on the left side (e.g. purple curve in Fig. \ref{fig:electrical_pickup}a), and 0.2 mV on the right side. The signal is relatively slowly varying (10s of~$\mu$s period) and nearly equal on each channel (varying by up to~$\sim 10\%$). Therefore, in analysis we can eliminate~$\sim 90\%$~of this spurious pick-up signal by subtracting the voltages measured on channel 0, i.e. the farthest infrared channel. On the right side, channel 0 is dominated by noise and/or spurious pick-up in all cases. On the left side, channel 0 is disconnected from the MA-PMT assembly. After collecting the data for this study, channel 0 of the was also disconnected from the right side MA-PMT assembly for simplicity. An example of the raw data before and after subtraction is shown in Fig. \ref{fig:electrical_pickup}. The spurious signal seems to come from imperfect grounding; by reducing the lengths of each of four key grounding wires from~$\sim 1$ meter to~$\sim$~20 cm, we reduced the pickup amplitude~$\sim 10$-fold to achieve the results shown here. Specifically, we reduced the lengths of 18 AWG braided wires from left and right oscilloscope grounding pins to left and right MA-PMT assembles, and from MA-PMT assemblies to the optical table.

In practice, this pickup problem is only a cause for concern at low temperatures, high currents, and durations longer than~$\sim 5$~$\mu$s, which we use for ambient pressure tests, but not for high pressure experiments. In other words, spurious signal is maximized in our proof-of-concept ambient pressure measurements. Nevertheless, improved design in future MA-PMT-assembly based systems could simplify data collection and analysis. To aid in future designs, note that we observe electrical pickup to be independent of MA-PMT supply voltage in the range 0 V to -900 V and approximately proportional to the amplitude of the Joule heating current pulse. To give intuition, we note that 1.5 to 3 mV pickup corresponds to~$\sim 1$~$\mu$A current across the 2.5 k$\Omega$ shunt resistors, a current that is 50 million times smaller than the Joule heating current. In other words, the magnitude of the pickup signal can be explained by 1 out of every 50 million Joule heating electrons traveling from each group of shorted PMT anodes, through a shunt resistor, the common ground, and through the Joule heated sample.

\section{Interpolation and discretization of the system response}
\label{SM:W_interpolation}
We interpolate the tungsten lamp spectrum,~$V_\textrm{W}$, by differentiating the cubic spline of its integral:

\begin{equation}
V_{\textrm{W,interp}} =  \frac{d}{d\lambda}\left(\textrm{Spline}\left(\int{V_\textrm{W}d\lambda}\right)\right)
\end{equation}
The result is a continuous function that preserves the integral of voltage in each wavelength band (i.e.~$V_{\textrm{W,interp}}d\lambda =  V_\textrm{W} \delta\lambda$ where~$\delta\lambda$ is the bandwidth). We use the ``interpolate.CubicSpline'' function with the option ``bc\_type = natural'' within the scipy library (version 1.9.1) in python. An example of the system response matrix is shown in Fig. \ref{fig:QE}. 

In practice, we discretize~$\lambda$~in steps of~$\sim 0.8$~nm, thereby increasing the density of interpolated wavelengths 100-fold compared to the measured wavelength. To make sure that the numerical implementation is reasonable, we compare best-fit temperatures for different discretizations: 100-fold, 10-fold, 4-fold, 3-fold, 2-fold, and 1-fold increase in~$\lambda$-resolution. First, we confirm that the ``1-fold'' case  gives the identical result to a simple python code that directly calculates the Planck fit by the using ``curve\_fit'' to find the best fit temperature and emissivity for the standard equation for spectroradiometry:
\begin{equation}
\epsilon B(T) =  \frac{V_\sam}{V_{\textrm{W}} } B(T = 2255 \textrm{ K})
\label{eqn:standard_planck}
\end{equation}
Second, we confirm that at temperatures within 200 K of the calibration temperature of the tungsten lamp, the correction due to interpolation (rather than use of equation \ref{eqn:standard_planck}) is in the range 10-30 K, because~$\frac{V_{\textrm{W,interp}}}{ B(\textrm{2255 K},\lambda)}$ is approximately constant with respect to wavelength. For example, the corrections for the platinum melting data and alumina melting data presented here are typically -20 K and +10 K, respectively. Third, we note that at temperatures~$\sim 1000$~K higher than the calibration temperature, a mere 3-fold increase is the density of interpolated values of~$\lambda$ is sufficient to achieve within 10 K of the result for 100-fold increase in~$\lambda$-resolution. Note that this suggests that a MA-PMT-based detector with~$\sim 3$-fold increase in spectral resolution would be sufficient to achieve~$\sim 20$ K precision without interpolation.

\bibliography{PMT}


\end{document}